\documentclass{aa}

\usepackage[english]{babel}
\usepackage{graphicx}
\usepackage{newlfont}
\usepackage{amssymb}
\usepackage{subfig}

\usepackage{amsmath}
\usepackage{latexsym}
\usepackage{amsthm}
\usepackage{natbib}
\usepackage{microtype}
\bibpunct{(}{)}{;}{a}{}{,} 

\usepackage[usenames,dvipsnames]{color}
\usepackage[bookmarks, colorlinks, breaklinks]{hyperref}  
\hypersetup{linkcolor=blue,citecolor=MidnightBlue,filecolor=black,urlcolor=MidnightBlue}

\usepackage{siunitx}
\DeclareSIUnit\parsec{pc}
\DeclareSIUnit\erg{erg}
\DeclareSIUnit\deg{deg^{-2}}
\DeclareSIUnit\count{cnt}
\DeclareSIUnit\solarmass{M_{\odot}}

\usepackage[varg]{txfonts}

\newcommand{\seciac}[1]{Sect. \ref{#1}}
\newcommand{\figiac}[1]{Fig. \ref{#1}}
\newcommand{\inv}[1]{\phantom{#1}}

\newfont{\gwpfont}{cmssq8 scaled 1000}
\newcommand{\rexcess}{{\gwpfont REXCESS}}

\def\plck{PLCK\,G266.6$+$27.3}

\def\YX {Y_{\textrm X}}

\def\TX {T_{\textrm X}}

\def\Rv {R_{500}}

\def\MY {$M_{500}$--$Y_{\textrm X}$}

\def\xmm{XMM-{\it Newton}}
\def\planck{{\it Planck}}
\def\chandra{{\it Chandra}}
\usepackage{color}

\begin{document}

\title{Resolving galaxy cluster gas properties at  $z\sim1$ \\
with \xmm\ and \chandra\ \\}

\author{I. Bartalucci\inst1, M. Arnaud\inst1, G.W. Pratt\inst1, J. D\'emocl\`es\inst1 \and R.F.J. van der Burg\inst1\and P. Mazzotta\inst{2,3}}

\institute{Laboratoire AIM, IRFU Service d’Astrophysique – CEA DRF – CNRS – Université Paris Diderot, B\^{a}t. 709, CEA-Saclay, 91191 Gif-sur-Yvette Cedex, France
           \and Dipartimento di Fisica, Università di Roma Tor Vergata, via della Ricerca Scientifica 1, 00133, Roma, Italy
           \and Harvard-Smithsonian centre for Astrophysics, 60 Garden Street, Cambridge, MA 02138, USA
          }
\date{Received August 10, 2016. Accepted October 7, 2016.}

\abstract{Massive, high-redshift, galaxy clusters are useful laboratories to test cosmological models and to probe structure formation and evolution, but observations are challenging due to cosmological dimming and angular distance effects. Here we present a pilot X-ray study of the five most massive ($M_{500}>5 \times 10^{14}$ M$_{\odot}$), distant ($z \sim 1$), clusters detected via the Sunyaev-Zel'Dovich effect. We optimally combine \xmm\ and \chandra\ X-ray observations by leveraging the throughput of \xmm\ to obtain spatially-resolved spectroscopy, and the spatial resolution of \chandra\  to probe the bright inner parts and to detect embedded point sources. Capitalising on the excellent agreement in flux-related measurements, we present a new method to derive the density profiles, which are constrained in the  centre by \chandra\ and in the outskirts by \xmm. 
We show that the \chandra-\xmm\ combination is fundamental for morphological analysis at these redshifts, the \chandra\ resolution being required to remove point source contamination, and the \xmm\ sensitivity allowing higher significance detection of faint substructures. Measuring the morphology using images from both instruments, we found that the sample is dominated by dynamically disturbed objects. 
We use the combined \chandra-\xmm\ density profiles and spatially-resolved temperature profiles to investigate thermodynamic quantities including entropy and pressure. From comparison of the scaled profiles with the local \rexcess\ sample, we find no significant departure from standard self-similar evolution, within the dispersion, at any radius, except for the entropy beyond 0.7~$\Rv$. The baryon mass fraction tends towards the cosmic value, with a weaker dependence on mass than that observed in the local Universe.  We make a comparison with the predictions from numerical simulations. The present pilot study demonstrates the utility and feasibility of spatially-resolved analysis of individual objects at high-redshift through the combination of \xmm\ and \chandra\ observations. Observations of a larger sample will allow a fuller statistical analysis to be undertaken, in particular of the intrinsic scatter in the structural and scaling properties of the cluster population.}
 \titlerunning{Study of gas properties in high-redshift galaxy clusters}
\authorrunning{Bartalucci et al.}

\maketitle

\section{Introduction}\label{sec:introduction}

High-mass, high-redshift galaxy clusters are of particular interest for a number of reasons. High-mass clusters are the ultimate examples of gravitational collapse, and so their evolution affords a unique probe of this process over cosmic time. Their abundance as a function of redshift is sensitive to the total matter content of the universe and its evolution, and their baryon fractions can be used as a distance indicator (\citealt{sasaki1996}, \citealt{pen1997}, \citealt{allen2004}, and \citealt{mantz2014} for a recent application).

Clusters emit in the X-ray band via the thermal emission of the hot and rarefied plasma in the intracluster medium (ICM). This emission can be used to measure the density, temperature, and heavy-element abundances of the gas, properties that are fundamental for the characterisation of the plasma thermodynamics. Global properties integrated over the cluster extent, such as temperature $T_\textrm{X}$ and the X-ray analogue of the Sunyaev-Zel'Dovich (SZ, \citealt{sunyaev1980}) signal $Y_\textrm{X} = M_\textrm{gas} \times T_\textrm{X}$ \citep{krav2006} can be used as proxies for the total mass. The analysis of radial profiles allows us to obtain measurements of the distribution of other fundamental thermodynamic quantities such as pressure and entropy. With the further assumption of hydrostatic equilibrium, the radial density and temperature distributions can be used to measure the total mass profile. 

X-ray observations, though, are particularly challenging for high-redshift galaxy clusters. The X-ray flux suffers from cosmological dimming, $S_\textrm{X} \propto (1+z)^{-4}$, limiting the photon statistics. The challenge is even more pronounced in the cluster outskirts because of the steep density gradient in these regions. Furthermore, the small apparent size of clusters at high redshift (typically $\sim 1-2$ arcmin radius), requires instruments with high spatial resolution to study the distribution of the emitting plasma. 

Deep X-ray observations of high-mass, high-redshift objects are scarce in the literature. These objects are rare and thus difficult to find; furthermore, they are difficult to observe because of their intrinsic faintness in the X-ray band \citep[see e.g.][for example]{ros09,san12,toz15}. An alternative approach, pioneered by \citet{spt2014}, consists of stacking a large number of shallow observations in order to derive the redshift evolution of the mean profiles. However, information on an individual cluster-by-cluster basis is necessarily lost in the stacking process.  The determination of individual profiles allows the average profile to be determined and also, crucially, the dispersion about it, thus linking the deviation from the mean to the (thermo-)dynamical history.   This approach provides essential information on cluster physics (see e.g. the analysis of local \rexcess\ entropy profiles by \citealt{pratt2010}), and is a key element for understanding the selection function of any survey. It is now clear that this function must be fully mastered in order to understand the properties of the underlying population, and for cosmological applications \citep[e.g.][]{angulo2012}. The selection function depends not only on global properties (i.e. the link between the observable and the mass and its dispersion), but also on the profile properties (e.g. more peaked clusters will have a higher luminosity; current SZ detection methods generally assume a certain pressure profile shape). 

Here we present a new method to combine observations obtained with two current-generation X-ray observatories. \xmm\ has the largest X-ray telescope effective area, which ensures sufficient photon statistics which helps to counterbalance the cosmological dimming. In turn, the high angular resolution of \chandra\ allows us to probe the inner parts of the cluster and to disentangle any point source emission from the extended cluster emission.
Using this method, we characterise the ICM of the five most massive ($M_{500}>5 \times 10^{14}\, M_{\odot})$\footnote{$M_{500}$ being the total mass of the cluster enclosed within $R_{500}$, which is the radius where the mean interior density is $500$ times the critical density of the Universe.}, distant ($z>0.9$) clusters currently known. They have been detected via the SZ effect in the South Pole Telescope (SPT, \citealt{spt1} and \citealt{spt2}) and \planck\  
surveys (\citealt{planckesz}, \citealt{planckpsz1}, and \citealt{planck_psz2})\footnote{There are no clusters in this mass and redshift range in the Atacama Cosmology Telescope (ACT) survey \citep{marriage2011}.}. For each object, we are able to make a quantitative measurement of the gas morphology and to determine the radial profiles for density, temperature, and related quantities (gas mass, pressure, and entropy). 

Combining observations from the two observatories allows us to efficiently probe both the inner regions and the outskirts. The observation depth was optimised to be able to measure the temperature in an annulus around $R_{500}$, allowing us to quantitatively study the radial scatter out to this distance. We investigate evolution through comparison with the results based on the X-ray selected Representative \xmm\ Cluster Structure Survey (\rexcess, \citealt{bohringer2007}); in particular, we compare to the average and $1\sigma$ dispersion of the  \rexcess\ density, temperature, pressure, and entropy profiles \citep{pratt2007,croston2008,arnaud2010,pratt2010}. Our sample being SZ-selected, the comparison with \rexcess\ also allows us to address the question of X-ray versus SZ-selection in the high-mass, high-redshift regime. For a fair comparison with SZ-selected samples, we also compare to the stacked results of \citet{spt2014}.  Finally, to contrast with theory, we also compare our sample with the five most massive z=$1$ clusters from the cosmo-OWLS numerical simulations of  \citet{amandine2014}. 
 
The paper is organised as follows. In \seciac{sec:data_preparation} we describe the \chandra\ and \xmm\ sample used in our work and the procedures to clean and process the datasets. In \seciac{sec:data_analysis} we present the analysis techniques used to perform the combination of the two instruments and to obtain radial profiles of density and 3D temperature, as well as the centroid shift used as dynamical indicator. In \seciac{sec:evolution} we address the question of evolution in the profiles through comparison with \rexcess, and discuss the question of evolution. We  compare our results with the high-redshift stacked entropy and pressure profiles of \citet{spt2014} and with the results of the numerical simulations of  \citet{amandine2014} in \seciac{sec:compxvp}  and \seciac{sec:compsimul}, respectively. We then draw our conclusions in \seciac{sec:conclusions}.

We adopt a flat $\Lambda$-cold dark matter cosmology with $\Omega_{M}=0.3$, $\Omega_{\Lambda}=0.7$, $H_{0} = \SI{70}{\kilo\meter \per\mega\parsec \per\second }$ and $h(z) = \sqrt{\Omega_{m} (1 + z)^{3} + \Omega_{\Lambda}}$. All errors are shown at the 68 percent confidence ($1\sigma$) level. All the fitting procedures are performed using $\chi^{2}$ minimisation.
\begin{figure}[!ht]
\begin{center}
\includegraphics[width=0.48\textwidth]{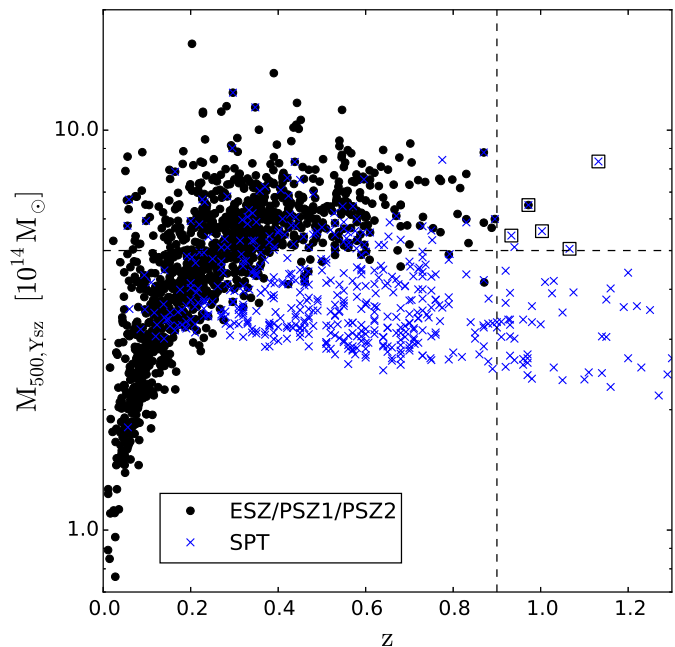} 
\end{center}
\caption{\footnotesize{Distribution of the galaxy clusters detected with \planck\ and SPT, denoted with black points and blue crosses, respectively. Dashed lines indicate the selection criteria defining our region of interest. Empty boxes highlight the $5$ galaxy clusters forming our sample. }}
\label{fig:dist_ammassi}
\end{figure}
\begin{table*}[t]
\caption{\footnotesize{The sample of the five galaxy clusters used in this work }}\label{tab:sample}
\begin{center}
\resizebox{\textwidth}{!} {
\begin{tabular}{cccccccccc}
\hline        
\hline
        Cluster name             & X-ray peak   							  					   & BCG           							  					    & $N_{H} ^{a}$                                        &  Redshift  & XMM exp.$^b$                      								& \chandra\ exp.$^c$      & $\langle w \rangle$       & BCG X-ray peak     \\ 
                                          &  $[J2000]$     							  					   & $[J2000]$    							  					    & $[10^{20}\si{\centi\metre}^{-3}]$   &                 & [\si{\kilo\second}]             						     & [\si{\kilo\second}]    & $[10^{-2} R_{500}]$   & distance   \\	
       &                                   &                       							  					   &                   							  					    &                                                      &                 M$1$ M$2$ PN   								&    ACIS                        &         XMM  \chandra\       &    $[10^{-2} R_{500}]$  \\
\hline \\[1.7mm]

SPT-CLJ 2146-4632 & $21$ $46$ $34.78 \inv{0} -46$ $32$ $54.06$   & $21$ $46$ $34.57 \inv{0} -46$ $32$ $57.20$  &  $1.64$                                        & $0.933$  & $147.5$        $157.4$           $102.2$          & $\inv{0}70.9$            &    $2.49 \pm 0.17 \inv{0} 2.46 \pm 0.78$  &    7.04       \\[1.7mm]
PLCKG266.6-27.3$^d$    & $06$ $15$ $51.83 \inv{0} -57$ $46$ $46.58$  & $06$ $15$ $51.77 \inv{0} -57$ $46$ $48.61$  &  $4.32$                                        & $0.972$  & $\inv{0}11.4$ $\inv{0}12.4$   $\inv{00}2.9$  & $226.7$                    & $0.68 \pm 0.18 \inv{0} 0.94 \pm 0.11$	     & 1.62      \\[1.7mm]
SPT-CLJ 2341-5119   & $23$ $41$ $12.23 \inv{0} -51$ $19$ $43.05$  & $23$ $41$ $12.37 \inv{0} -51$ $19$ $44.62$  &  $1.21$                                        & $1.003$  & $\inv{0}86.9$ $\inv{0}93.6$   $\inv{0}44.5$ & $\inv{0}77.8$            & $1.96 \pm 0.16 \inv{0} 2.98 \pm 0.87$      & 2.06     \\[1.7mm]
SPT-CLJ 0546-5345   & $05$ $46$ $37.22 \inv{0} -53$ $45$ $34.43$  & $05$ $46$ $37.63 \inv{0} -53$ $45$ $30.49$  &  $6.79$                                        & $1.066$  & $126.0$         $127.9$          $112.9$          & $\inv{0}67.9$            &  $1.16 \pm 0.11 \inv{0} 0.96 \pm 0.34$    & 5.63       \\[1.7mm]
SPT-CLJ 2106-5845   & $21$ $06$ $05.28 \inv{0} -58$ $44$ $31.70$  & $21$ $06$ $04.60 \inv{0} -58$ $44$ $28.21$  &  $4.33$                    					  & $1.132$  & $\inv{0}26.1$ $\inv{0}25.9$   $\inv{0}16.1$ & $\inv{0}70.8$            & $1.55 \pm 0.21 \inv{0}  1.90 \pm 0.30$    & 6.08 \\[1.7mm]
\hline
\end{tabular}
}
\end{center}
\footnotesize{Notes: $^a$ The neutral hydrogen column density absorption along the line of sight is determined from the LAB survey \citep{kalberla2005}. $^{b,c}$ We report the total exposure time after cleaning procedures. $^d$ SPT name: SPT-CLJ 0615-5746.}

\end{table*}
\section{Data preparation}\label{sec:data_preparation}
\subsection{Sample}\label{subsec:sample}

Figure \ref{fig:dist_ammassi} shows all the confirmed clusters detected by the SPT and \textit{Planck} surveys in the $M$--$z$ plane. There are five clusters with $M_{500} > 5 \times 10^{14}$ M$_{\odot}$ at $z>0.9$, all of which were detected by SPT, and one, \plck, was detected by both surveys. These are shown by black boxes in the figure.

We obtained an \xmm\ Large Programme to observe four of these clusters in AO-13 (proposal $074440$, PI M. Arnaud). 
The exposure times (typically 100 ks before cleaning) were tuned so as to obtain sufficient counts to extract high-quality spatially resolved temperature profiles up to $R_{500}$.  The  fifth cluster,  \plck, had already been observed  to similar depth with \chandra\  ($\sim 225$ ks, proposal $13800663$, PI P. Mazzotta).  In addition to our deep \xmm\ data, the four SPT clusters  had previously been observed as part of the \chandra\ X-Ray Visionary Project proposal $13800883$ (P.I. B. Benson). Exposure times for these observations (typically 80 ks), were tuned so as to obtain $\sim 2\,000$ counts per cluster to allow for global properties to be measured \citep{spt2014}.  The original $\sim 25$ ks snapshot observation of  \plck, obtained as part of the \planck-\xmm\ validation programme, is described in \citet{planck_xmm_plckg266}. 

Observation details are given in Table~\ref{tab:sample}. All the \xmm\  observations were taken using the European Photon Imaging Camera (EPIC, \citealt{turner2001} and \citealt{struder2001}), combining the data taken with the MOS$1$, MOS$2$, and PN cameras. 
 \chandra\ observations were taken using the Advanced CCD imaging spectrometer (ACIS, \citealt{garmire2003}).  \xmm\ and \chandra\ images of all five objects are shown in Appendix~\ref{appx:a1}.

\begin{table*}[!ht]
\caption{\footnotesize{Global properties}}\label{tab:global_thermo}
\begin{center}
\scalebox{0.8}{
\begin{tabular}{cccccc}
\hline        
\hline   
Cluster	&	$T_\textrm{X}$ $^a$						   &		$\YX$ $^b$								        &		$R_{500}$ $^c$		                 &	$Mg_{500}$ $^d$	             &	    $M_{500,\YX}$	$^e$   \\	
			& [\si{\kilo\electronvolt}]  	   & $[10^{14} M_{\odot} \si{\kilo\electronvolt}]$ &	[\si{\kilo\parsec}]     		             &  $[10^{13} M_{\odot}]$      &  $[10^{14} M_{\odot}]$   \\
			&								        &	 											  			    &							                          &										&									\\
\hline \\[1.7mm]
 SPT-CLJ 2146-4632   & $4.80^{+0.24}_{-0.21}$   & $2.18^{+0.17}_{-0.16}$  							&   $726.97^{+9.77}_{-11.10}$        	& $4.54^{+0.13}_{-0.14}$    &	 $3.14^{+0.13}_{-0.14}$		\\[1.7mm]
 PLCKG266.6-27.3     & $11.04^{+0.56}_{-0.56}$ & $13.42^{+1.00}_{-1.04}$  							&   $1002.58^{+14.07}_{-14.38}$  	& $12.16^{+0.03}_{-0.03}$   &	 $8.61^{+0.37}_{-0.37}$ 		\\[1.7mm]
 SPT-CLJ 2341-5119    & $7.08^{+0.36}_{-0.36}$   & $3.72^{+0.27}_{-0.28}$ 							     &   $777.65^{+10.66}_{-10.89}$       & $5.26^{+0.12}_{-0.12}$     &	 $4.17^{+0.17}_{-0.17}$ 	\\[1.7mm]
 SPT-CLJ 0546-5345    & $7.68^{+0.38}_{-0.33}$   & $4.24^{+0.26}_{-0.30}$								&   $773.96^{+10.02}_{-8.94}$          & $5.52^{+0.11}_{-0.10}$     &	 $4.41^{+0.17}_{-0.15}$		\\[1.7mm]
 SPT-CLJ 2106-5845    & $10.04^{+0.83}_{-0.86}$  & $10.08^{+1.10}_{-1.16}$ 							    &   $882.94^{+18.20}_{-18.77}$   	& $10.10^{+0.31}_{-0.29}$ 	 &	 $7.07^{+0.45}_{-0.42}$		\\[1.7mm]
\hline 
\end{tabular}}
\end{center}
\footnotesize{Notes:  $^a$ Spectroscopic temperature measured in the $[0.15-0.75]\,R_{500}$ region. $^b$ $\YX$ is the product of  $M_\textrm{gas,500}$ and $\TX$  \citep{krav2006}. $^d$ The gas mass within $R_{500}$, derived using the combined density profiles. $^{c,e}$ $R_{500}$ and  $M_{500,\YX}$  are determined iteratively using the \MY\ relation of \citet{arnaud2010}, assuming self-similar evolution.}
\end{table*}

\subsection{Data preparation}\label{subsec:data_prep}
We processed \chandra\ observations using the \chandra\ Interactive Analysis of Observations (CIAO, \citealt{fruscione2006}) ver. $4.7$ and the calibration database\footnote{cxc.harvard.edu/caldb} ver. $4.6.5$. 
The latest version of calibration files were applied following the prescriptions detailed in \verb?cxc.harvard.edu/ciao/guides/acis_data.html? using the \verb?chandra_repro? tool. 
To process the \xmm\ datasets, we used the Science Analysis System\footnote{cosmos.esa.int/web/xmm-newton} pipeline ver. $14.0$ and calibration files as available in December $2015$. New event files with the latest calibration applied were produced using the \verb?emchain? and \verb?epchain? tools.

\subsection{Data filtering}\label{subsec:data_filtering}
We reduced the contamination from high-energy particles using the Very Faint Mode status bit\footnote{cxc.harvard.edu/cal/Acis/Cal\_prods/vfbkgrnd} and the KEYWORD pattern for the \chandra\ and \xmm\ datasets, respectively. In particular, we removed from the analysis all the events for which the keyword PATTERN is $>4$ and $>13$ for MOS$1,2$, and PN cameras.
To remove periods of anomalous count rate, i.e. flares, we followed the prescriptions described in Markevitch's COOKBOOK\footnote{cxc.harvard/contrib/maxim} and \citet{pratt2007} for \chandra\ and \xmm, respectively. For both datasets, we extracted a light curve for each observation and removed from the analysis the time intervals where the count rate exceeds $3\sigma$ times the mean value. 
If there were multiple observations of the same objects, these were merged after the processing and cleaning procedures. We list in Table~\ref{tab:sample} the effective exposure times, after cleaning, for each instrument.

Point sources were identified using the CIAO \texttt{wavdetect} tool \citep{freeman2002} on $[0.5-2], [2-8]$ and $[0.5-8]$\,\si{\kilo\electronvolt} exposure-corrected images. 
For \xmm\ we ran Multiresolution wavelet software \citep{sta98} on the exposure-corrected $[0.3-2]$\, \si{\kilo\electronvolt} and $[2-5]$ keV images. We then inspected by eye each list to check for false detections and missed point sources. Within $\SI{3}{\arcmin}$ of the aimpoint of each \xmm\ observation, we used the \chandra\ point source list as reference. The positions of point sources were compared and, in the case of missed or confused sources, we defined a circular region of  $\SI{15}{\arcsecond}$ radius to be used as mask. 

\subsection{Background estimation}\label{sec:background}
X-ray background emission can be separated into sky and instrumental components. The latter is the result of high-energy particles interacting with the detectors and with the telescope itself. To evaluate this component for \xmm, we used filter wheel closed \citep{kuntz2008} datasets. For \chandra, we generated mock datasets using the analytical particle background model described in \citet{bartalucci2014}. 
We normalised these datasets based on the total count rates measured over the entire field of view in the $[9.5-10.6]$\, \si{\kilo\electronvolt} band for ACIS and $[10-12],[12-14]$\, \si{\kilo\electronvolt} for the MOS$1$-MOS$2$ and PN cameras, respectively. We then skycast the normalised instrumental datasets to match our observations and applied the same point source masking.
 In addition to the instrumental component, we also produced datasets reproducing out-of-time (OOT) events, following the prescriptions of \citet{chandra_back} and using the SAS-\verb?epchain? tool for \chandra\ and \xmm. After skycasting and point source removal, we merged the OOT and instrumental datasets. From now on, we refer to this merge as ``instrumental background datasets''. 
 
The sky background is due to a local component, formed by the Local Hot Bubble and the halo Galactic emission, and an extragalactic component (see, e.g., \citealt{snowden1995} and \citealt{kuntzsnowden2000}). The latter is the result of the superimposition of unresolved point sources, namely the cosmic X-ray background (CXB, \citealt{giacconi2001}). 
In \seciac{subsec:sx_profiles} and \seciac{subsec:kt_profiles} below we describe in detail how we estimated and subtracted these components  for imaging and for spectroscopic analysis.

\subsection{Vignetting correction}
We corrected for the vignetting effect following the method described in \citet{arnaud2001}, where we assigned a WEIGHT keyword to each detected event. This is defined as the ratio of the effective area at the aimpoint to the area at the event position (in detector coordinates) and energy. By doing so, the WEIGHT represents the effective number of photons we would detect if the instrument had the same response as at the aimpoint.
To compute the WEIGHT we used the SAS \verb?evigweight? tool and the procedures described in \citet{bar16}  for \xmm\ and \chandra, respectively. For consistency, when subtracting the instrumental background, we also computed the WEIGHTs for the background datasets.

\section{Radial profiles}\label{sec:data_analysis}

All the techniques described in this section are applied to both the \chandra\ and \xmm\ datasets, unless stated otherwise.  Global cluster parameters are  estimated self-consistently within $R_{500}$ via iteration about the \MY\ relation of \citet{arnaud2010}, assuming self-similar evolution.
The quantity $\YX$ is defined as the product of $M_\textrm{g,500}$, the gas mass within $R_{500}$, and $\TX$, the spectroscopic temperature measured in the $[0.15-0.75]\, R_{500}$ aperture.  The X-ray properties of the clusters and resulting refined $\YX$ values are listed in Table~\ref{tab:global_thermo}.

\subsection{X-ray peak and BCG positions}
To assess how the choice of the centre affects our profiles, we used both the X-ray peak position and the brightest cluster galaxy (BCG) location as the centre for surface brightness and temperature profile extraction. Because of the higher spatial resolution, we determined the X-ray peak using \chandra\ images in the $[0.5-2.5]\, \si{\kilo\electronvolt}$ band, smoothed using a Gaussian whose width ranges from three to five arcseconds.  We determined the BCG positions on Spitzer/IRAC data taken from the archive (PID:60099, PID:70053, and PID:80012). If available (for all clusters except SPT-CLJ 2146-4632), positions were refined using archival HST imaging in the F814W band, which is part of HST programs $12246$ and $12477$. We give the positions in Table~\ref{tab:sample} and they are shown in the right column in \figiac{fig:gallery} by black and white crosses for the X-ray peak and BCG position, respectively.

\subsection{Surface brightness profiles}\label{subsec:sx_profiles}
We extracted vignetting-corrected and instrumental-background-subtracted surface brightness profiles from \xmm\ datasets using concentric annuli in the $[0.3-2]\, \si{\kilo\electronvolt}$ band, each annulus being $\SI{3.3}{\arcsec}$ wide. For \chandra\ datasets we extracted the profiles in the $[0.7-2.5]\, \si{\kilo\electronvolt}$ band, each annulus being $\SI{2}{\arcsec}$ wide.
We evaluated the sky background component in a region free from cluster emission, i.e. where the instrumental background-subtracted surface brightness profiles flatten. 
Sky background subtracted surface brightness profiles were then binned to have a significance of at least $3\sigma$ per bin.
\subsection{Density profiles}\label{sec:density_profiles}
Density profiles were determined by applying the deprojection with the regularisation technique described in \citet{croston2006}. Briefly, from the surface brightness profiles we produced PSF-corrected and deprojected emission profiles. 
For \xmm\ we used the PSF parametrisation described in \citet{ghizzardi2001}, while for \chandra\ we assumed a perfect PSF and only account for geometrical deprojection. We converted the emission measure profiles to gas density using a factor that depends on temperature and redshift, namely $\lambda(T,z)$. 
As already pointed by several works (e.g. \citealt{martino2014} and \citealt{schellenberger2015}), there is a mismatch between temperatures measured by \chandra\ and \xmm: on average, \xmm\ temperatures are $\sim 15\%$ lower than those of \chandra\ at 10\, keV.
However, as discussed in \citet{bar16}, the conversion factor is only weakly dependent on the temperature so that the offset between the two instruments is negligible for the computation of the density profiles. The conversion factors for \chandra\ and \xmm\ were computed using their respective temperature profiles and assuming an average abundance of 0.3 $Z_{\odot}$. If we did not have a temperature profile (see below), we used the global temperature $\TX$ listed in Table~\ref{tab:global_thermo} to obtain the conversion factor. Using a conversion factor evaluated via a spatially resolved temperature or a single global value results in a negligible difference in the final density profile\footnote{Temperature profiles typically vary by $\sim 30\%$ \citep{vikhlinin2006,pratt2007,arnaud2010}. However, these variations induce only $\sim 1\%$ effects on the resulting density profile, thus any difference in the radial temperature profile as measured by \xmm\ and \chandra\ \citep[e.g.][]{donahue2014} would have a similarly negligible effect.}.
 
\begin{figure*}[!ht]
\begin{center}
\includegraphics[width=0.95\textwidth]{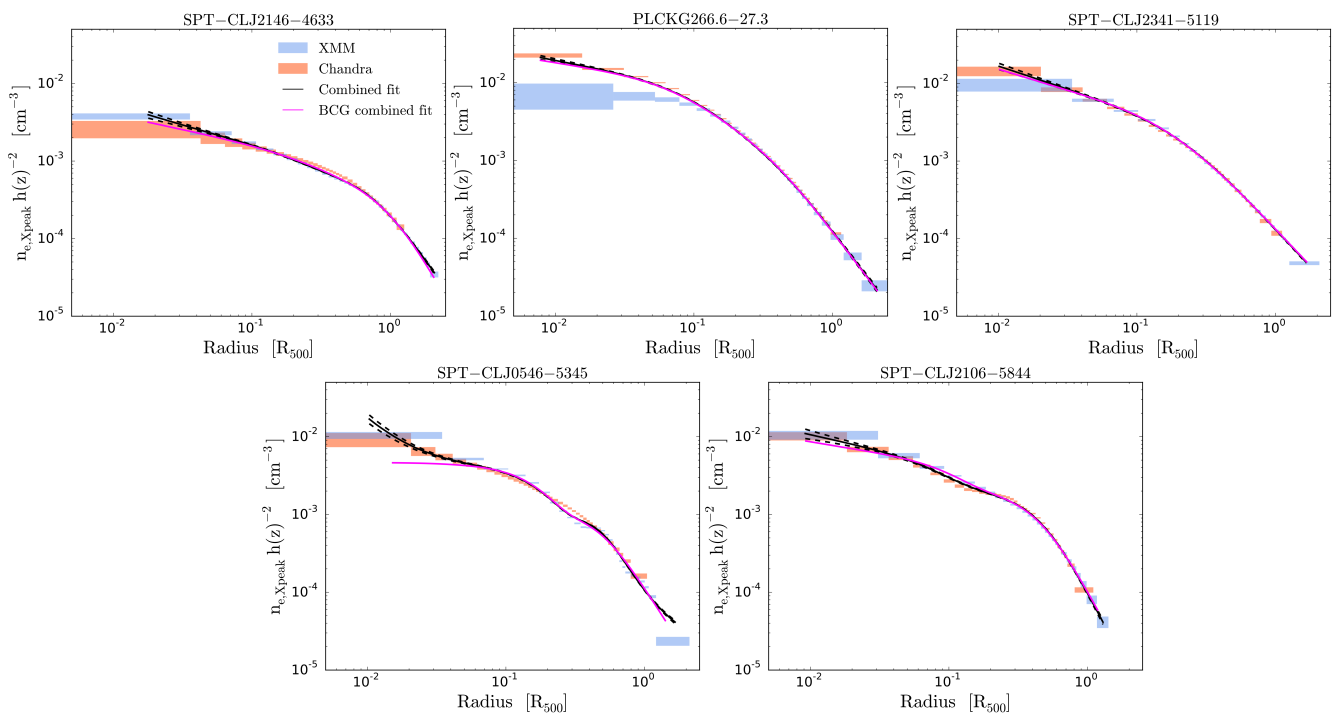}  
  \end{center}
 \caption{\footnotesize{Normalised, scaled, deprojected, density profiles centred on the X-ray peak. Blue and orange rectangles represent \xmm\ and \chandra\ datasets, respectively. The solid black line shows the combined density profile resulting from the simultaneous fit to the \xmm\ and \chandra\ data, as discussed in the text. Its uncertainties are shown with the dashed line. The magenta line shows the simultaneous fit for the profiles centred on the BCG. }}
 \label{fig:hyb_density_xpeak}
\end{figure*}

Figure \ref{fig:hyb_density_xpeak} shows the deprojected density profiles computed from \xmm\ and \chandra\ with blue and orange rectangles, respectively, using the X-ray peak as centre. 
The profiles are scaled by $h(z)^{-2}$ to account for self-similar evolution \citep{croston2008}. The corresponding BCG-centred profiles are shown in \figiac{fig:hyb_density_bcg}. 

\begin{figure*}[!ht]
 \begin{center}
\includegraphics[width=0.85\textwidth]{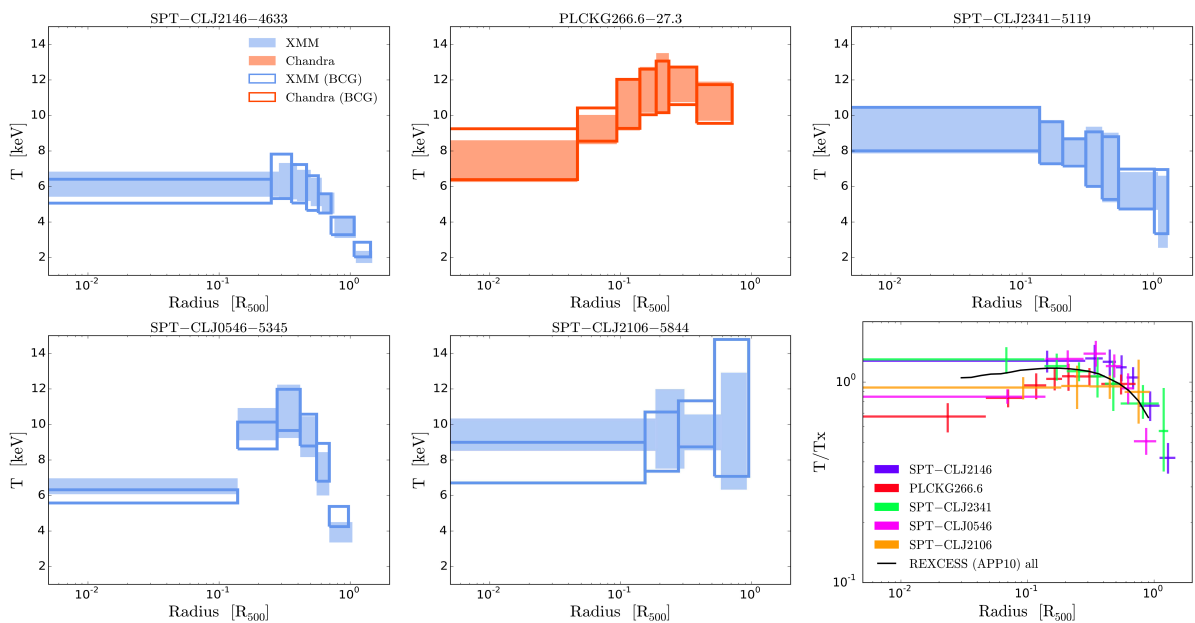} 
  \end{center}
 \caption{\footnotesize{First five panels: 3D temperature profiles. Blue and orange rectangles are profiles determined using \xmm\ and \chandra\ datasets, respectively. Filled and empty boxes represent profiles computed using the X-ray peak and the BCG as centre, respectively. Bottom right panel: 3D X-ray peak-centred temperature profiles scaled to their global $\TX$ values. The solid black line shows the average value of the \rexcess\ temperature profiles (\citealt{pratt2007,arnaud2010}, APP10). Profiles are colour coded according to the mass estimated from $M_{500,\YX}$, the most massive being red and least massive being blue.}}
 \label{fig:kt_plots}
\end{figure*} 

\subsection{Temperature profiles}\label{subsec:kt_profiles}
\subsubsection{Temperature profile extraction}
To measure the projected 2D temperature profiles, we analysed the spectra extracted from concentric annuli centred on the X-ray peak, each bin width being defined to have at least a signal-to-noise ratio of $30\sigma$ above the background level in the $[0.3-2]$ keV and in the $[0.7-2.5]$ keV band for \xmm\ and \chandra, respectively. Spectra were binned to have at least $25$ counts per energy bin after instrumental background subtraction. Following these prescriptions, we were able to define at least four annuli in all \xmm\ datasets except for \plck,  where the deep \chandra\ observation allowed us to define seven annuli. For the other four clusters, the \chandra\ observations were not deep enough to determine a temperature radial profile. For this reason, from now on all the temperature-based quantities for \plck\ are computed using the \chandra\ dataset, while for the others we use the \xmm\ observations.

Our spectroscopic analysis consists of a spectral fit using a combination of models accounting for the cluster and sky background emission.
We modelled the latter using two absorbed MEKAL models \citep{mewe1985, mewe1986, kaastra1992, liedhal1995} plus an absorbed power law with index fixed to $1.42$ (\citealt{lumb2002}) to account for the Galactic and the CXB emission, respectively. For one of the MEKAL models, the absorption was fixed to $0.7 \times 10^{20}\si{\centi\meter}^{-3}$; for the other two models the absorption was fixed to the Galactic value along the line of sight, as listed in Table~\ref{tab:sample}. The absorption was accounted for in the fit by using the WABS model (\citealt{morrison1983}).
  To estimate the sky background normalisations and temperatures, the model was fitted to a spectrum that was extracted from a region that is free of cluster emission, vignetting-corrected, and instrumental-background-subtracted. Once determined, we fixed these sky background models and simply scaled them by the ratio of the areas of the extraction regions.  
  We modelled the cluster emission using an absorbed MEKAL whose absorption was fixed to the values given in Table~\ref{tab:sample}. From the fit of this component we then determined the normalisation and the temperature. The photon statistics of the sample did not allow us to measure the abundances with an error lower than $30\%$, so we fixed the abundance to $0.3 Z_{\odot}$.
Fitting was undertaken in XSPEC\footnote{heasarc.gsfc.nasa.gov/xanadu/xspec/} using the $[0.3-11]\, \si{\kilo\electronvolt}$ and $[0.7-10]\, \si{\kilo\electronvolt}$ range for \xmm\ and \chandra, respectively. To avoid prominent line contamination in the \xmm\ data, we excluded the $ [1.4-1.6]\, \si{\kilo\electronvolt}$ spectral range for all three camera datasets  and the $[7.45-9.0]\, \si{\kilo\electronvolt}$ band for PN only \citep[see e.g.][]{pratt2007}.
 We convolved all the models by the appropriate response matrix file (RMF), which were computed using the SAS \verb?rmfgen? and the CIAO \verb?mkacisrmf? tools. Convolved models were then multiplied by the ancillary response file computed at the aimpoint using the SAS \verb?arfgen? and the CIAO \verb?mkarf? tools. 

\subsubsection{3D Temperature profiles}
We determined the 3D temperature profiles by fitting a parametric model similar to that described in \citet{vikhlinin2006} to the 2D profiles. The models were convolved with a response matrix that simultaneously takes into account projection and PSF redistribution (the latter being set to zero for the \chandra\ data). In convolving the models, the weighting scheme introduced by \citet{vikh_multit} (see also \citealt{mazzotta2004}) was used to correct for the bias introduced by fitting isothermal models to a multi-temperature plasma.  We computed the uncertainties via Monte Carlo simulations of $1000$ random Gaussian realisations of the projected temperature profiles.

Using a parametric model may over-constrain the resulting temperature profile and hence underestimate the error. For this reason, if the resulting error in a specific bin was smaller than the one in the 2D profiles we used the latter as the final error. The 3D \xmm\ and \chandra\ temperature profiles are shown with blue and orange rectangles, respectively, in \figiac{fig:kt_plots}. The empty boxes show the profiles obtained when centring the annuli on the BCG. The two profiles are consistent within the uncertainties for all objects. 
 \begin{figure}[!ht]
 \begin{center}
\includegraphics[width=0.43\textwidth]{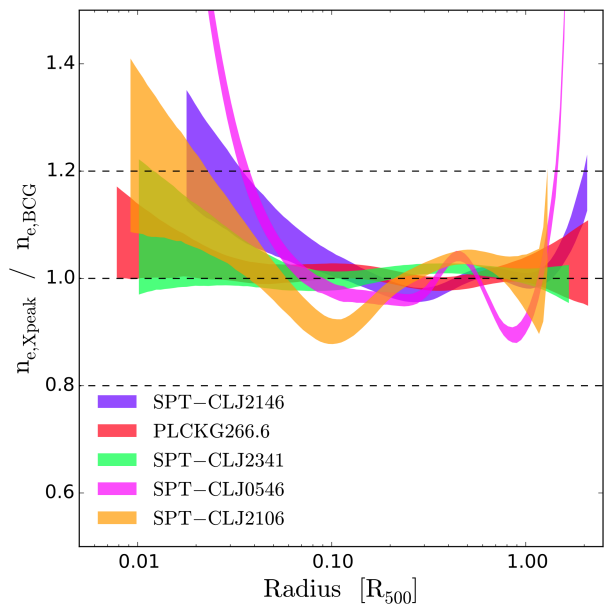}  
  \end{center}
 \caption{\footnotesize{Ratio between the combined density profiles centred on the X-ray peak and those centred on the BCG. The colour scheme is the same as in the bottom right panel of \figiac{fig:kt_plots}.}}
 \label{fig:bcg_vs_xpeak}
\end{figure}
\subsubsection{Effect of the PSF on $\TX$}

The value of $\TX$ derived from  the \xmm\  spectrum in the projected $[0.15$--$0.75]~\Rv$ radial range was not corrected for the PSF. At these redshifts, $0.15\, \Rv$ is about 15 arcseconds, comparable with the size of the PSF. Redistribution of photons from the cluster core to the aperture in question could bias $\TX$, although we expect the effect to be small, since the  temperature profiles are  rather flat in the cluster centre and no  density profile is particularly peaked.   To quantify the  effect, we computed the spectroscopic-like temperature in the  $[0.15$--$0.75]~\Rv$  aperture using the best-fitting convolved temperature profile model, with and without  taking into account the \xmm\ PSF.  We first  checked that the value derived  taking into account the  PSF, $T_\textrm{sl,psf}$,  is consistent with the directly measured $\TX$ values for each cluster within the error bars.  The  median ratio is 1.03.  We then compared this  spectroscopic-like temperature,  $T_\textrm{sl,psf}$,  with that obtained without taking into account the PSF,  i.e. the model value for a perfect instrument. The difference is 1\% on average. The effect of the PSF is indeed negligible on $\TX$.

\subsection{\chandra\ XMM combination}\label{sec:chandra_xmm_combination}
\begin{figure*}[t]
 \centering
 \begin{minipage}[c]{0.40\textwidth}
\resizebox{\textwidth}{!} {
\includegraphics[]{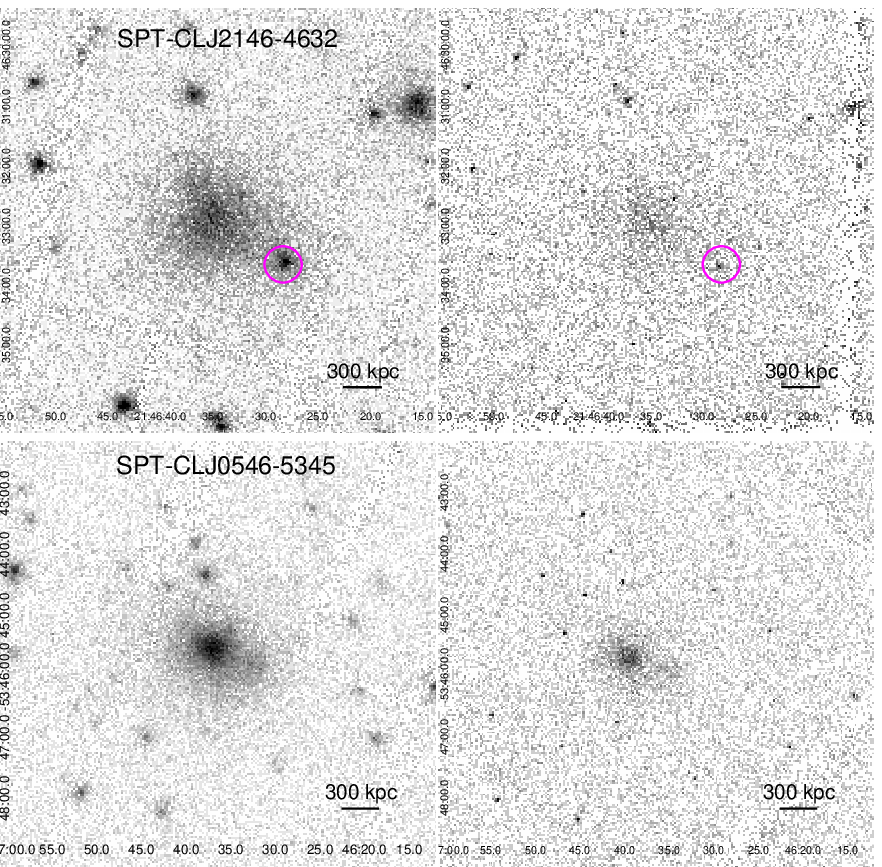}}
\end{minipage}
\hspace{0.1\textwidth}
 \begin{minipage}[c]{0.40\textwidth}
\resizebox{\textwidth}{!} {
\includegraphics[]{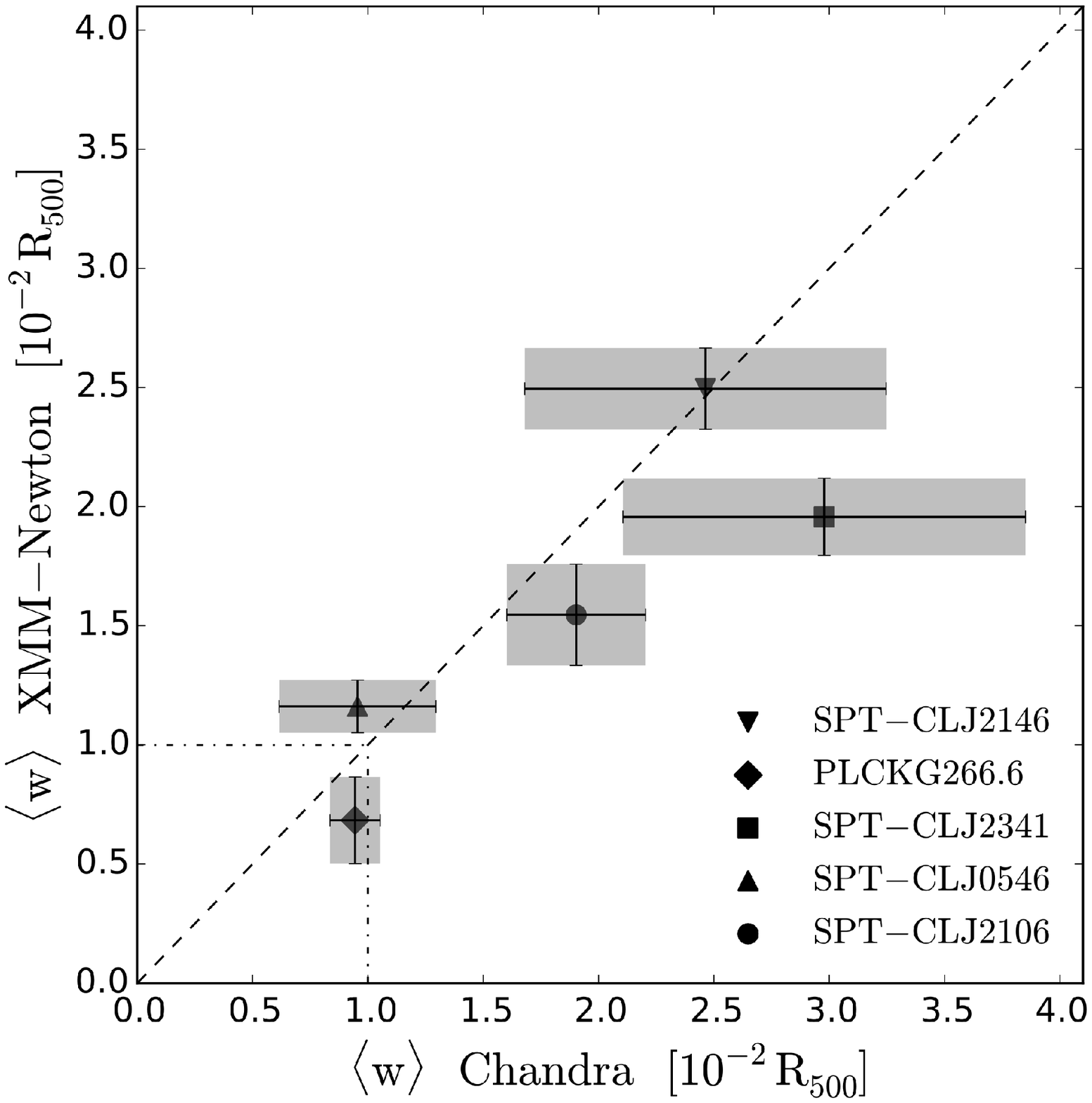}}
\end{minipage}
 \caption{\footnotesize{Left panel: \xmm\ (left) and \chandra\ (right) images of SPT-CLJ 2146-4632 (top) and SPT-CLJ 0546-5345 (bottom) in the $[0.3-2]$\, keV band. The images pixel size is $2\arcsec$, and for ease of comparison, the same scale is used for all. The magenta circle identifies the point source that contaminates the substructure emission in the \xmm\ image of SPT-CLJ 2146-4632. Right panel: centroid shift parameter, $\langle w \rangle$, values computed using \chandra\ and \xmm\ in units of $10^{-2}\, R_{500}$. The dashed line is the identity relation. The two dot-dashed lines define the region where the clusters are considered to be dynamically relaxed (see text).
}}
 \label{fig:morpho_cxo_vs_xmm}
\end{figure*}
As shown in \figiac{fig:hyb_density_xpeak} and \figiac{fig:hyb_density_bcg}, there is excellent agreement between the deprojected \chandra\ and \xmm\  density profiles beyond $\sim0.1R_{500}$. The two profiles contain complementary information. The higher effective area of XMM-\textit{Newon} constrains the density profiles up to $\sim1.5 R_{500}$ in all cases. Furthermore, the higher photon statistics are fundamental to clearly detect the presence of substructure, such as the``knee'' features in the density profiles of SPT-CLJ 0546-5345 and SPT-CLJ 2106-5845. Conversely, the higher \chandra\ resolution is useful to probe the inner regions, where the radial binning of the profile can be finer than that of \xmm.
We thus combined the two deprojected density profiles by undertaking a simultaneous fit with a parametric modified beta model similar to that described in \citet{vikhlinin2006}.  The resulting fit is a smooth and regular function, that can be easily integrated or differentiated, which at the same time efficiently combines information from the two instruments. As discussed in \citet{bar16}, \chandra\ and \xmm\ density profile shapes are in good agreement with a normalisation offset of the order of $1\%$. For this reason, during the simultaneous fit we added a normalisation factor accounting for this effect as a free parameter. 
Errors were computed by performing $1000$ Monte Carlo realisations of the simultaneous fit of the deprojected density profiles. The improvement due to the instrument combination is evident in \figiac{fig:hyb_density_xpeak} and \figiac{fig:hyb_density_bcg}, where we show the combined parametric density profile and associated errors with solid and dotted black lines, respectively. In all cases the resulting profiles are well constrained at large radii with the XMM-\textit{Newon} datasets, whereas in the central regions \chandra\ drives the fit. The combined profiles also retain information on the presence of features such as the knee observed in the \xmm\ profile of SPT-CLJ 0546-5345, shown in \figiac{fig:hyb_density_xpeak}.

The density profiles of \plck\ exhibit strong disagreement between the two instruments in the inner region, independent of whether the profile is centred on the X-ray peak or the BCG.  The \chandra\ image in the right column  of \figiac{fig:gallery} shows that the cluster emission in the core presents a complex structure. Taking the BCG as reference, there is a small region of low surface brightness emission to the north-west and a horseshoe-like region of high surface brightness emission around it to the north. The typical scale of these features is of $\sim \SI{4}{\arcsec}$, so that the different resolving power of the instruments can significantly change the apparent emission distribution in these inner regions. 

The accuracy of the combined profile in the inner parts allows us to study the impact of the choice of the centre.
We repeated the same analysis for the density profiles centred on the BCG position and show the combined profiles with a black line in \figiac{fig:hyb_density_bcg}. 
For comparison, BCG (X-ray-peak) centred profiles are shown in \figiac{fig:hyb_density_xpeak} (\figiac{fig:hyb_density_bcg}) with magenta lines. As expected, BCG-centred profiles are shallower in the inner parts, but exhibit similar behaviour to the X-ray peak-centred profiles beyond $0.1\,R_{500}$. 
Figure~\ref{fig:bcg_vs_xpeak} shows the ratio of X-ray peak- to BCG-centred profiles. As expected, the clusters for which the difference between the X-ray peak and the BCG positions are larger (notably SPT-CLJ 2146-4632, SPT-CLJ 0546-5345, and SPT-CLJ 2106-5845) exhibit more variation between profiles in the core. The choice of centre thus appears to affect the inner parts of the profiles, but does not seem to affect the outer part.

 In the following we use the combined density profiles to perform our study and to derive all the other quantities. 

\subsection{Morphological analysis}\label{sec:morpho_discussion}

\begin{figure*}[t]
 \centering
 \begin{minipage}[c]{0.40\textwidth}
\resizebox{\textwidth}{!} {
\includegraphics[]{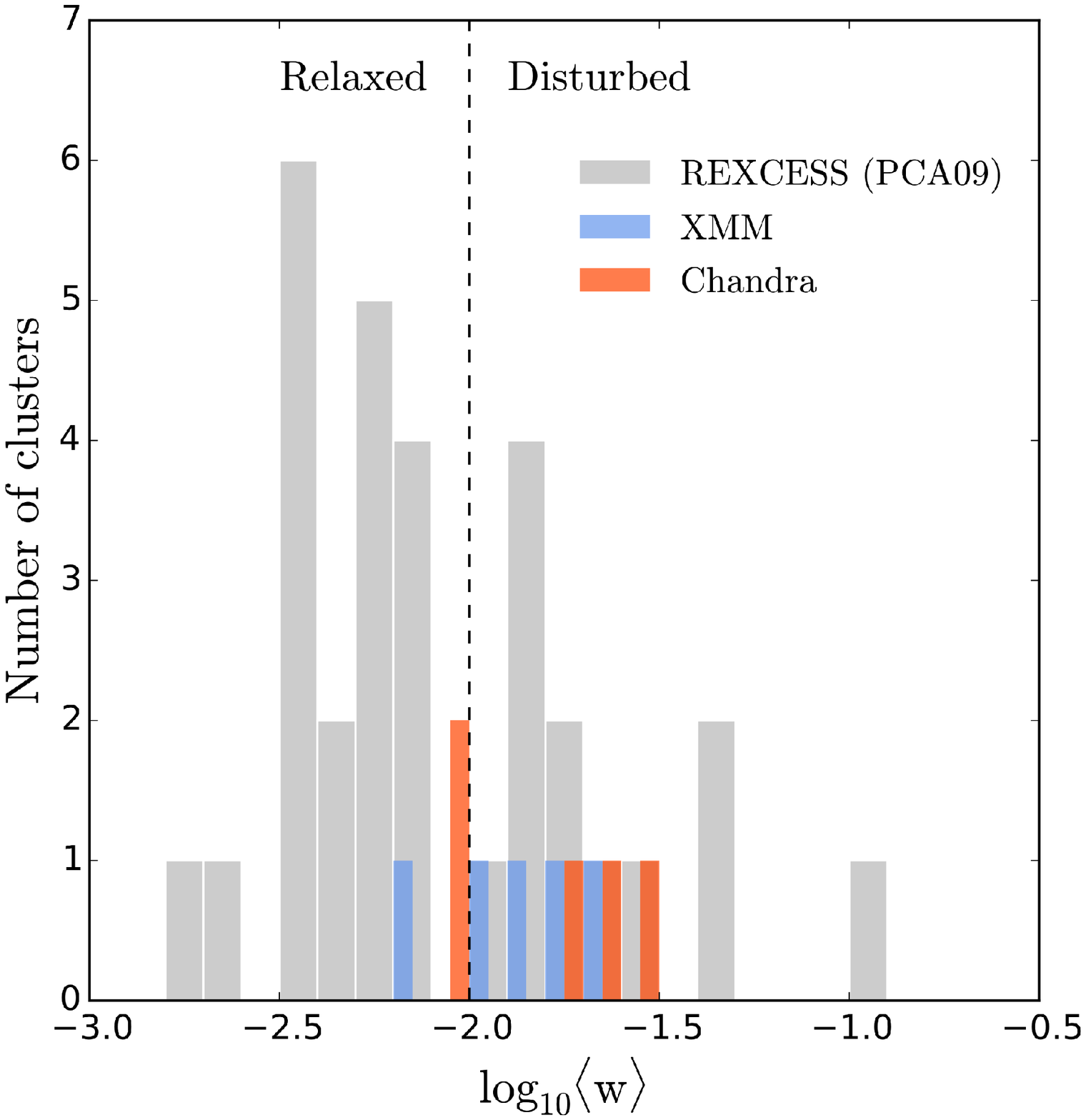}}
\end{minipage}
\hspace{0.1\textwidth}
 \begin{minipage}[c]{0.40\textwidth}
\resizebox{\textwidth}{!} {
\includegraphics[]{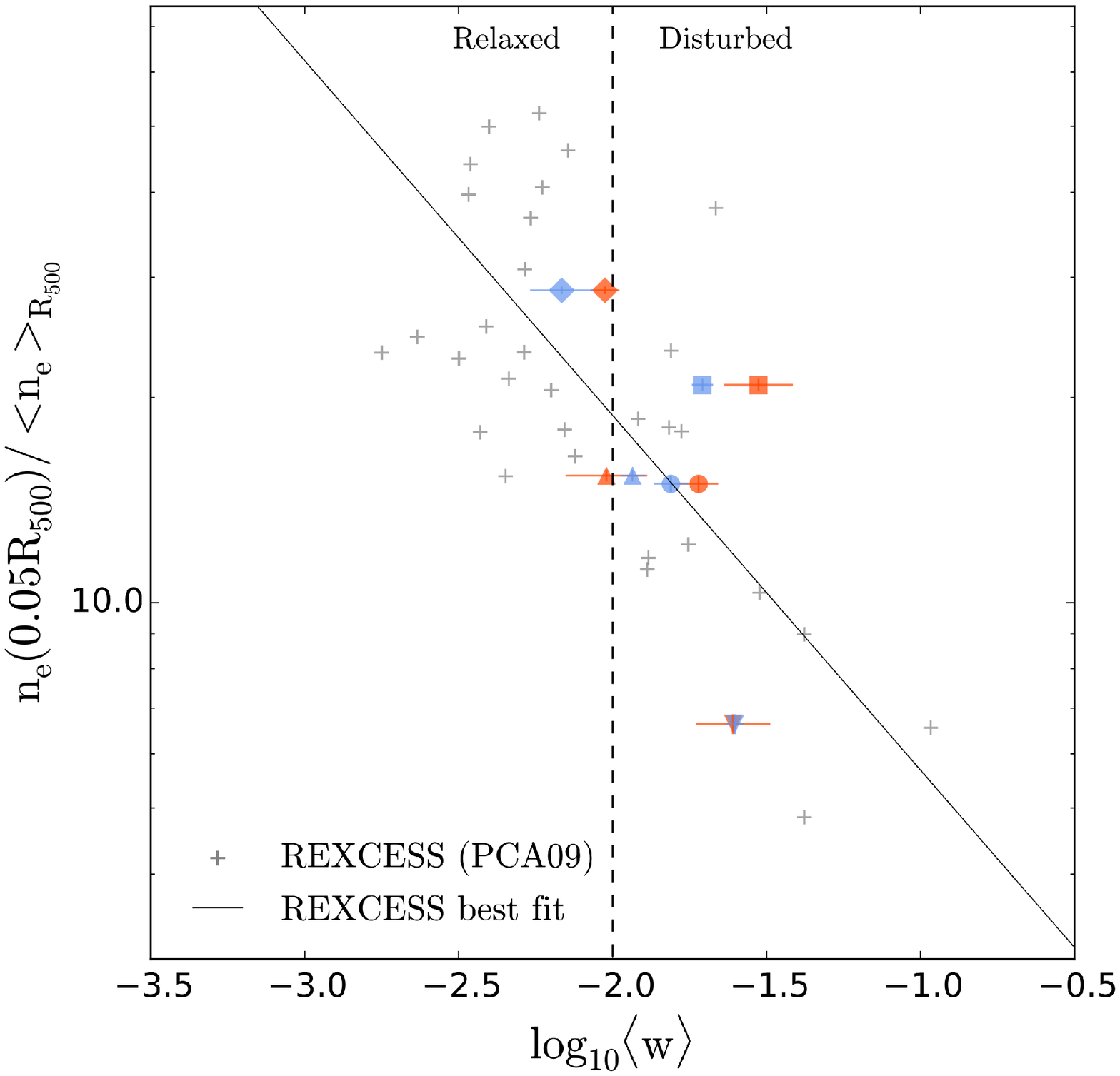}}
\end{minipage}
 \caption{\footnotesize{Left panel: the number of clusters as a function of $\langle w \rangle$ for the \rexcess\ sample (\citealt{pratt2009}, PCA09) are plotted in grey. Our objects are plotted in blue and red for \xmm\ and \chandra, respectively. The dashed line separates morphologically disturbed clusters from the rest. We plot \xmm\ and \chandra\ lines using a binsize of $0.05$ for clarity reasons, whereas for the histogram computation we use a binsize of $0.1$. 
Right panel: distribution of the scaled density computed at $0.05R_{500}$ as a function of $\langle w \rangle$. Grey crosses show the \rexcess\ sample, while blue and red points represent \xmm\ and \chandra\ measurements, respectively, of our sample. The black solid line is the best-fit of the \rexcess\ sample. The dashed line divides morphologically disturbed objects from the rest. }}
 \label{fig:w_mine_vs_rexcess}
\end{figure*}
We adopted the centroid shift value introduced by \citet{mohr1993}, namely $\langle w \rangle$, as an objective estimator of the morphological state of the cluster.  The centroid shift is defined as the standard deviation of the projected distance between the X-ray peak and the centroid, measured in concentric circular apertures. To compute $\langle w \rangle$ we followed the implementation described in \citet{maughan2008}. Briefly, the centroids were measured on exposure-corrected and background-subtracted images in the $[0.3-2]\, \si{\kilo\electronvolt}$ and $[0.5-2.5]\, \si{\kilo\electronvolt}$ bands, for \xmm\ and \chandra, respectively. To enhance the sensitivity to faint substructures, we masked the contribution in the inner region of the cluster and computed the centroid in ten concentric annuli in the range $[0.1-1]\,R_{500}$. 
Images were binned using the same pixel size of $\SI{2}{\arcsec}$. 
To test the dependence of $\langle w \rangle$ on image resolution, we repeated the same analysis on \chandra\ images binned with a pixel size of 1". We found consistent results within the $1\sigma$ error bar.
We computed the error by undertaking  $100$ Poisson realisations of the image and taking the $1\sigma$ standard deviation of the resulting $\langle w \rangle$ distribution. Values for $\langle w \rangle$ are listed in units of $10^{-2} \Rv$ in Table~\ref{tab:sample}. 

The right panel of \figiac{fig:morpho_cxo_vs_xmm} shows the comparison between the $\langle w \rangle$ as computed using \xmm\ and \chandra. Dot-dashed lines highlight the region of relaxed clusters, $\langle w \rangle <0.01$, as defined for the \rexcess\ sample \citep{pratt2009}. There is good agreement between the measurements, $\langle w \rangle$ being consistent at $1 \sigma$ in all cases. However, for a given exposure time,  the higher photon statistics of the \xmm\ observations yield typically $\sim 3$ times smaller uncertainties. This can be important for the classification of objects as relaxed or otherwise (see e.g. SPT-CLJ 0546-5345 in Fig.~\ref{fig:morpho_cxo_vs_xmm}). 

The left panel of \figiac{fig:morpho_cxo_vs_xmm} shows example  \xmm\ and \chandra\ images of SPT-CLJ 2146-4632 and SPT-CLJ 0546-5345 in the $[0.3-2]\, \si{\kilo\electronvolt}$ band. They have been background subtracted and corrected for exposure time and are shown on exactly the same scale.
Substructures are highlighted with blue dotted circles in the \xmm\ images in \figiac{fig:gallery}. In the case of SPT-CLJ 0546-5345 the substructure at $\sim 0.4\, R_{500}$ in the south-west sector is detected at much higher significance with \xmm\ allowing unambiguous classification of its dynamical state. \xmm\ clearly detects diffuse emission at $\sim 0.9 \, R_{500}$ in the south-west sector of SPT-CLJ 2146-4632, to which the $\langle w \rangle$ is sensitive. However, this emission also contains some contribution from a point source, highlighted with a magenta circle in \figiac{fig:morpho_cxo_vs_xmm}.  The \chandra\ image allows us to accurately determine the point source location, and hence mask it properly in the \xmm\ image. It is important to note that a complete morphological analysis of high-redshift clusters thus needs both instruments to efficiently detect faint diffuse emission and to account for point source contamination.

 \begin{figure*}[t]
 \centering
 \resizebox{0.9\textwidth}{!} {
\includegraphics[]{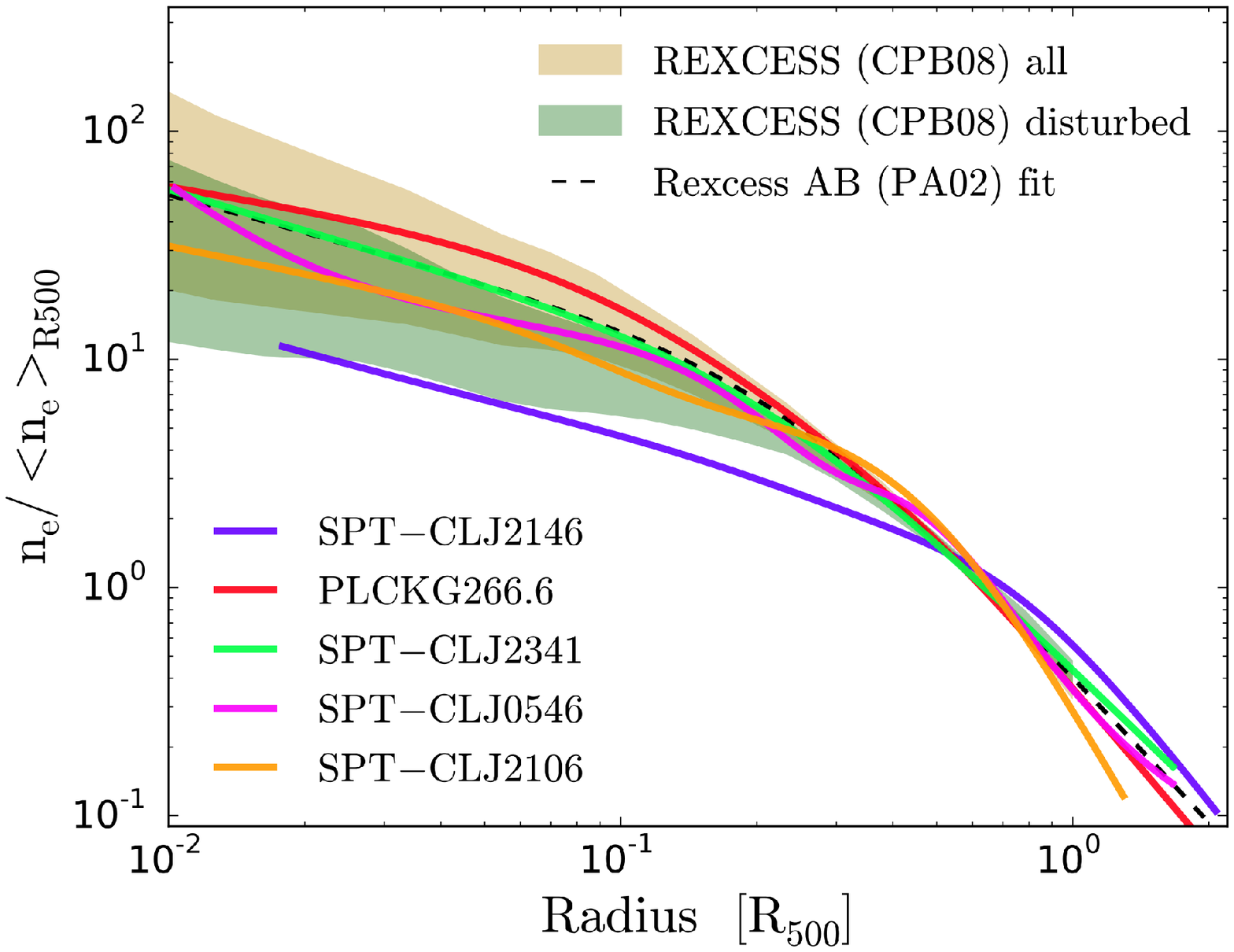}
\hspace{0.05\textwidth}
\includegraphics[]{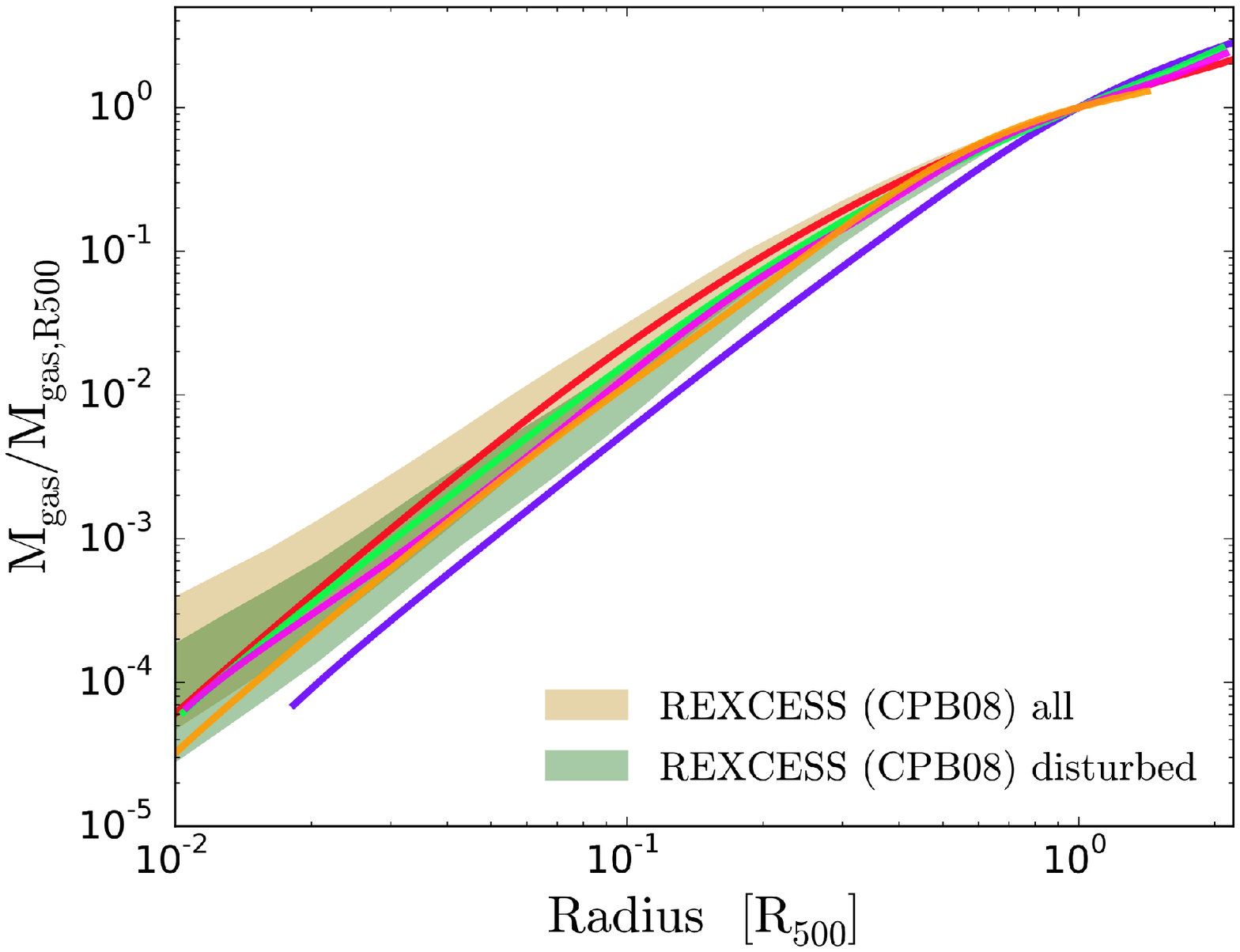}
}
  \caption{\footnotesize{Left panel: density profiles scaled by the integrated density within $R_{500}$. Gold and green shaded areas are the scatter computed from \rexcess\ density profiles (\citealt{croston2008}, CPB08) using the full sample and only the disturbed objects \citep{pratt2009}, respectively. The dashed line is the fit of the average \rexcess\ density profile with an AB model  (\citealt{pratt2002}, PA02). The profile colour-coding is the same as in the bottom right panel of \figiac{fig:kt_plots}. Right panel: gas mass profiles scaled to the average gas mass within $R_{500}$. The colour-code is same as in the left panel.}}
 \label{fig:density_mgas_profiles}
\end{figure*}

\section{Evolution of cluster properties}
\label{sec:evolution}

\subsection{Dynamical status}
Figure \ref{fig:w_mine_vs_rexcess} compares the $\langle w \rangle$ distribution of  \rexcess\ \citep{pratt2009} and our sample. Taking the centroid shift values from \xmm\ as a baseline, we find that our sample is dominated by disturbed clusters using the \rexcess\ definition ($\langle w \rangle > 0.01$). Only \plck\ appears relaxed, confirming the initial results of \citet{planck_xmm_plckg266}.  This result is expected from structure formation theory, in which the merging rate at high redshift is higher than in the local universe (see e.g. \citealt{gott2001}  or \citealt{hopkins2010}). 

In the following, we compare scaled profiles considering both the full \rexcess\ sample and the subsample composed only of the dynamically disturbed \rexcess\ clusters. 

\subsection{Scaled density and gas-mass profiles}
The global cluster properties computed from the \MY\ relation, assuming self-similar evolution, are given in Table \ref{tab:global_thermo},  together with the details of the computation. The left panel of \figiac{fig:density_mgas_profiles} shows the density profiles scaled to the average density integrated within $R_{500}$. The profiles are colour-coded based on their temperature $\TX$, as reported in Table~\ref{tab:global_thermo}, where blue is $\SI{4}{\kilo\electronvolt}$ and red is $\SI{14}{\kilo\electronvolt}$. We also show the $1 \sigma$ dispersion of the full \rexcess\ sample, and the subset of morphologically disturbed \rexcess\ clusters using golden and green envelopes, respectively. For a reference value, we fit the full \rexcess\ sample using an AB model (see \citealt{pratt2002}), and this is shown with a black dotted line.
   The profiles are, on average, in good agreement within the \rexcess\ scatter, but below the average reference value. This is not the case considering only disturbed clusters, where our sample is in excellent agreement within the scatter. 
The right panel of \figiac{fig:w_mine_vs_rexcess} shows the scaled density computed at $0.05R_{500}$ as a function of $\langle w \rangle$. The centroid shifts computed using \chandra\  and \xmm\ are plotted in red and blue, respectively. The majority of our sample has a large $\langle w \rangle$ and a flatter central density, and thus these objects lie in the parameter space of disturbed clusters.
 We suggest that there is no evident sign of evolution in the density profile shape, apart from that related to the  evolution of the dynamical state. The profile of SPT-CLJ2146-4632, shown with the blue line in \figiac{fig:density_mgas_profiles}, lies outside the \rexcess\ $1\sigma$ envelope. We are not able to asses whether the difference is due to a particular behaviour, for instance related to its (relatively) very cool temperature, or if it is truly an outlier. Interestingly, we also identify a similar outlier in the simulated dataset (see below).

From the density profiles we compute the gas-mass profiles by integrating in spherical shells. These are shown in \figiac{fig:density_mgas_profiles}, where the dimensionless gas-mass profiles are scaled to the value integrated within $R_{500}$. There is good agreement between our sample and \rexcess. Furthermore, as discussed in \cite{croston2008}, we find the same behaviour of the gas mass as a function of temperature: hotter clusters have a larger gas fraction within $R_{500}$.

We compute the same quantities using the profiles centred on the BCG and we find the same results.

\subsection{Baryon content}

\begin{figure}[]
 \begin{center}
\includegraphics[width=0.45\textwidth]{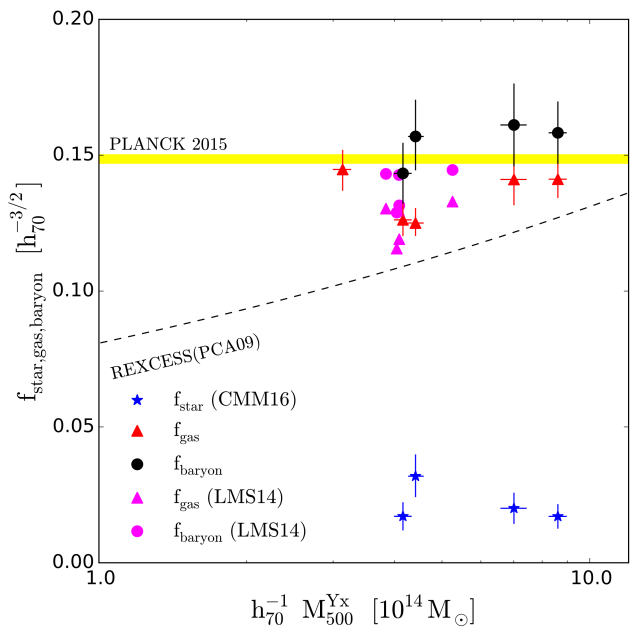} 
  \end{center}
 \caption{\footnotesize{Fraction of mass in stars, gas and total baryons at $R_{500}$. Blue stars and red triangles represent the stellar mass fraction (\citealt{chiu2016}, CMM16) and the gas-mass fraction at $R_{500}$, respectively. The black points represent the sum of the two (i.e. the total baryon mass). For SPT-CLJ2 146-4632 there are no stellar mass fraction measures available. Magenta points show the gas and total baryon fraction of five clusters with $M_{500} > 4.5 \times 10^{14}$ M$_{\odot}$ from the cosmo-OWLs simulations (\citealt{amandine2014}, LMS14).The yellow area is the baryon fraction as measured from \citet{planck2015}. The dashed black line is the variation of gas-mass fraction with mass in the \rexcess\ sample (\citealt{pratt2009}, PCA09).}}
 \label{fig:gas_fraction}
\end{figure}

One of the open issues related to the mass content of galaxy clusters is the evolution of the baryon fraction, defined as the ratio of stellar plus ICM content to the total cluster mass. To estimate the total baryon fraction for our sample we use the stellar mass computed by \citet{chiu2016}, who computed the stellar mass for 14 SPT clusters, four of which are in common with our sample. (The stellar mass is not available for SPT-CLJ 2146-4632). Figure \ref{fig:gas_fraction} shows the stellar, gas, and total baryonic mass fraction at $R_{500}$. We also plot the best-fit \rexcess\  gas mass fraction versus total mass relation \citep{pratt2009}. While our sample is small and the scatter of the \rexcess\ fit is large, the gas-mass fractions of the four clusters in the present sample lie above the local reference; in addition, there does not appear to be a variation of gas or baryon fraction with mass. The total baryon fractions of the four clusters for which we have measurements are all in agreement with the cosmic baryon fraction measured by  \citet{planck2015}.

\subsection{Thermodynamical  profiles}
Pressure and entropy profiles are computed using the density and temperature profiles described in the preceding sections. Since the radial sampling of the density profiles is finer, we interpolate them to match the radial binning of the temperature profiles. Typically we have five temperature profile bins, but in the following plots we show solid continuous lines to guide the eye.

 \begin{figure*}[tbp]
\centering
\begin{minipage}[c]{0.45\textwidth}
\resizebox{\hsize}{!} {
\includegraphics[]{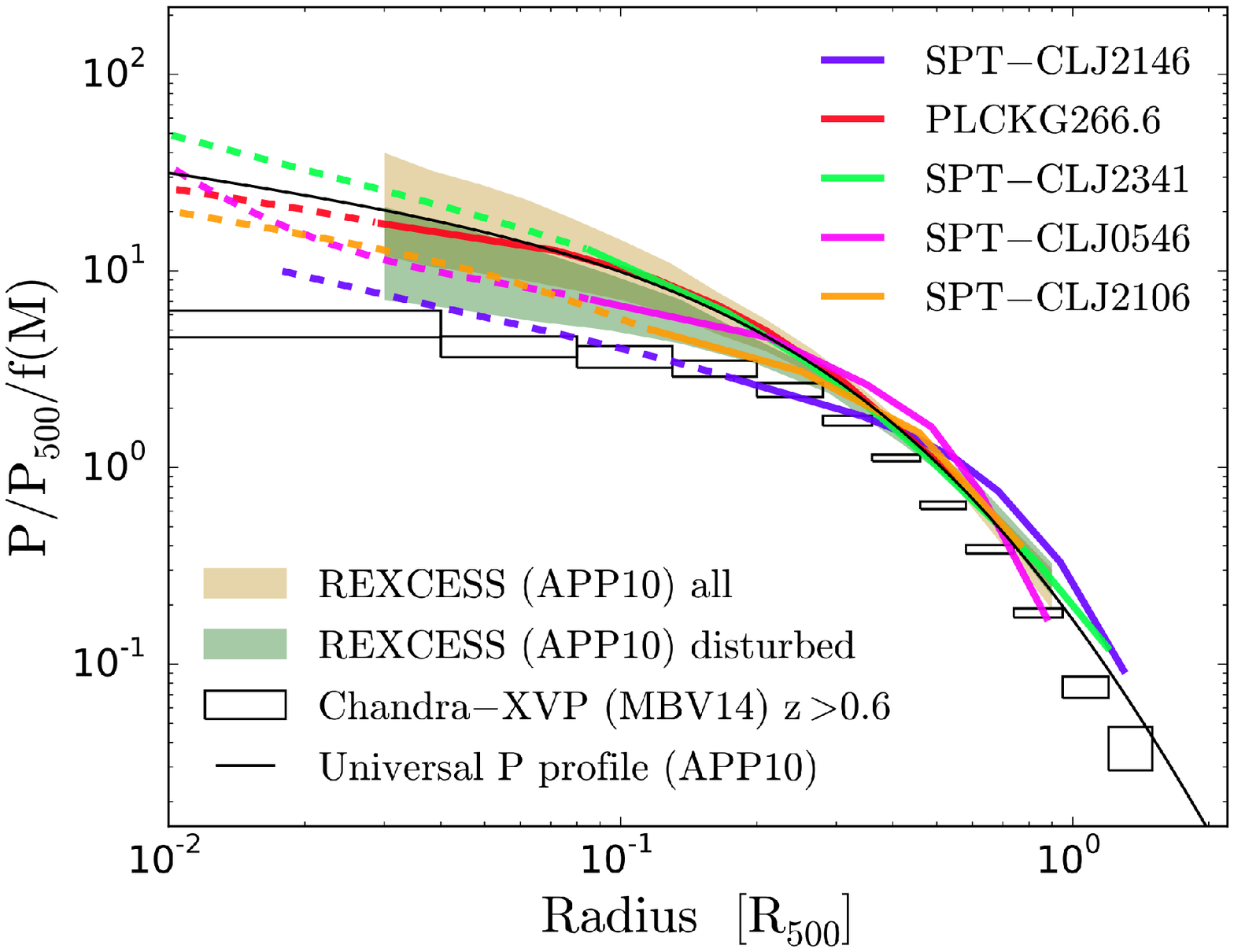}}
\end{minipage}
\hspace{0.1\textwidth}
\begin{minipage}[c]{0.35\textwidth}
\resizebox{\hsize}{!} {
\includegraphics[]{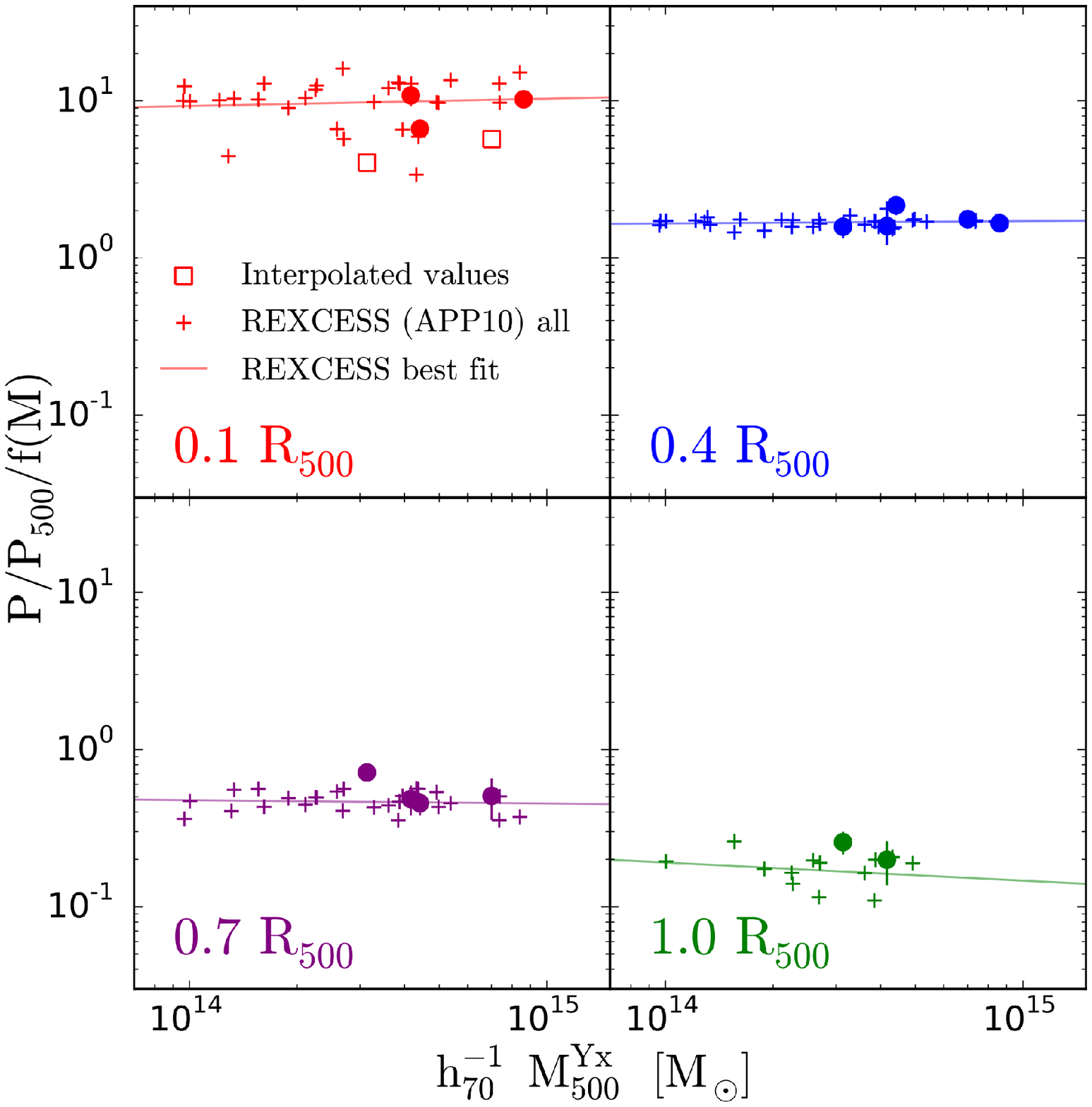}} 
\end{minipage}
\\[2mm]
\begin{minipage}[c]{0.45\textwidth}
\resizebox{\hsize}{!} {
\includegraphics[]{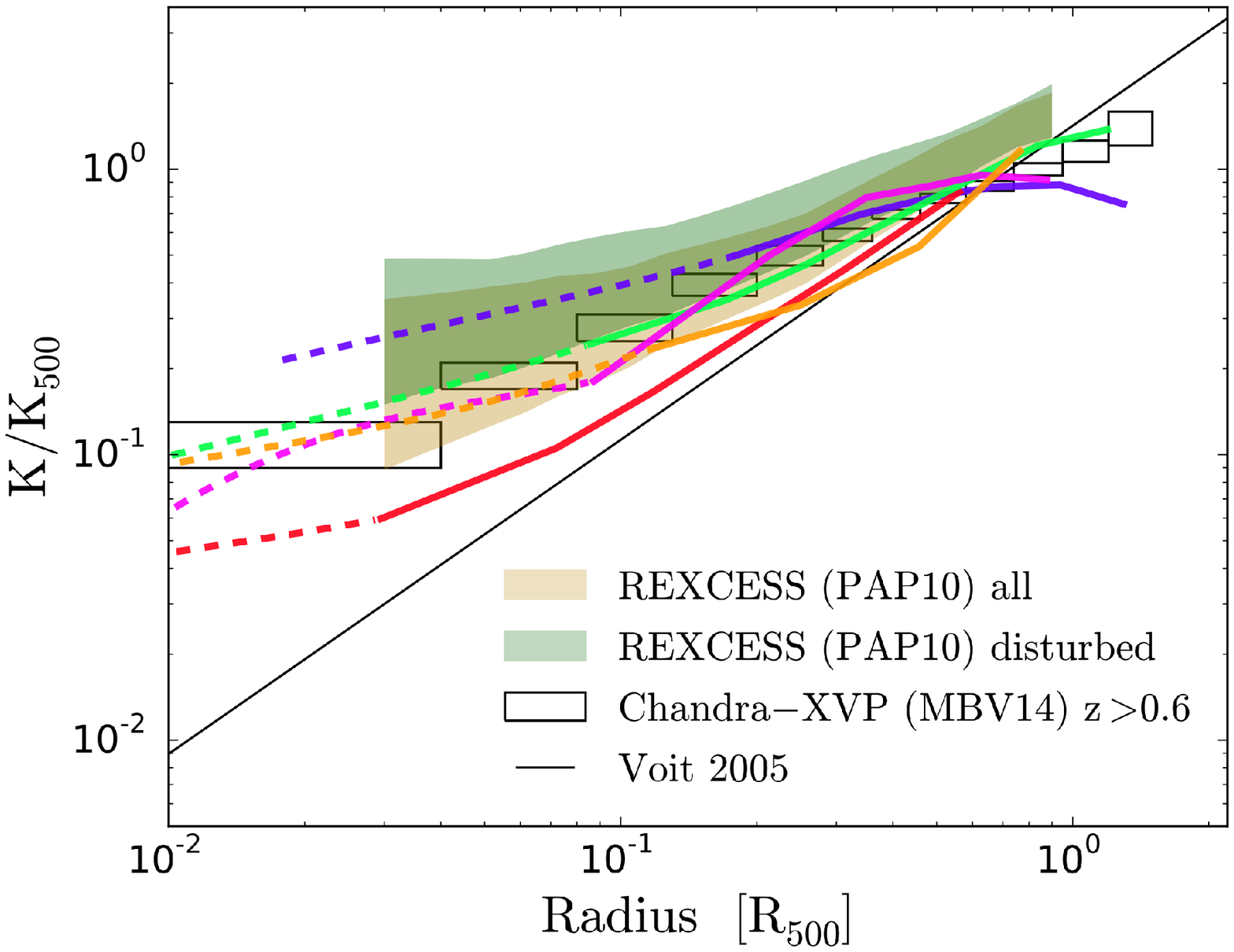}} 
\end{minipage}
\hspace{0.1\textwidth}
\begin{minipage}[c]{0.35\textwidth}
\resizebox{\hsize}{!} {
\includegraphics[]{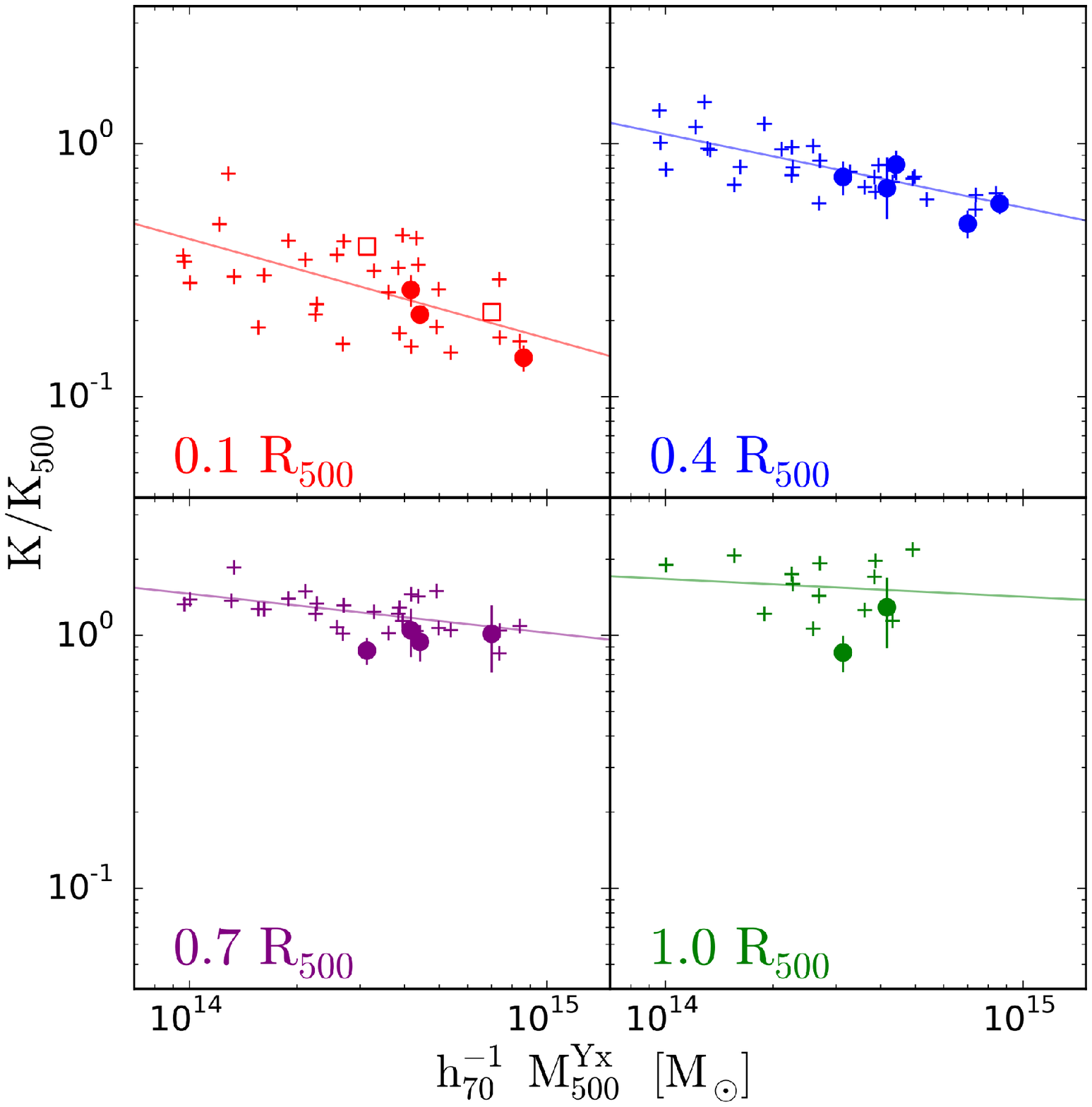}}
\end{minipage}

 \caption{\footnotesize{Top left: dimensionless pressure profiles scaled by $P_{500}$ and $f(M)$, colour-coded as in the bottom right panel of \figiac{fig:kt_plots}. The $1 \sigma$ dispersion in the \rexcess\ pressure profiles (\citealt{arnaud2010}, APP10) considering the full sample and the disturbed susbset are plotted using the same colour-code as in the left panel of \figiac{fig:density_mgas_profiles}. The black solid line is the universal pressure profile from \citet{arnaud2010}. Black boxes show the stacked pressure profile of \citet{spt2014} (MBV14) and its uncertainty. Top right: scaled pressure values computed at fixed radii as a function of $M_{500}$ of our sample (points) and \rexcess\ (crosses). 
 The black solid line shows the best power-law fit to the \rexcess\ sample. To avoid confusion, the \rexcess\ errors are not plotted. 
Bottom left and right: same but for the entropy profiles, scaled by $K_{500}$ from \citet{pratt2010} (PAP10). The black solid line shows the  theoretical entropy profiles from the gravity-only simulations of \citet{voit2005all}.}}
 \label{fig:pressure_entropy_disp}
\end{figure*}

\subsubsection{Pressure profiles}

As discussed in \citet{arnaud2010}, pressure profiles typically exhibit a smaller dispersion than the other quantities because of the anti-correlation between density and temperature. The left panel of \figiac{fig:pressure_entropy_disp} shows the pressure profiles
scaled by $P_{500}$, and by the factor $f(M) = (M_{500}/3 \times  10^{14} h_{70}^{-1}  M_{\odot})^{0.12}$  to account for the mass dependency (see equation $10$ of \citealt{arnaud2010}). 
We also show the $1\sigma$ \rexcess\ envelopes for the full sample and the disturbed subsample, and the universal pressure profile from \citet{arnaud2010}.
To better visualise the behaviour in the central parts, we also show the pressure profiles extrapolated assuming a flat temperature profile in the core. 
Our scaled profiles are in excellent agreement with \rexcess\ over the entire radial range, considering both the full sample and the disturbed subsample. Interestingly, there is a very small scatter above $\sim 0.2 R_{500}$.
As with the density, the pressure profile of SPT-CLJ2146-4632 deviates significantly  from the others, lying outside the \rexcess\ envelopes.

To investigate how the properties of our sample depends on the mass, we show  in the right panel of \figiac{fig:pressure_entropy_disp} the pressure values computed at fixed radii compared to \rexcess. For each fixed radius, we fit the \rexcess\ sample with a power-law function of the form $Y=(N/N_{0})^{\alpha}$. We show the result of the fit with a black solid line. In the inner radius panel at $0.1R_{500}$, we show the interpolated pressure profile with hollow squares. There is excellent agreement with the \rexcess\ best fit at $0.1, 0.4$ and $0.7\,R_{500}$; in addition, the average deviation of the two samples is in excellent agreement. At $R_{500}$ we have only two points, so we are not able to tell if there is a real offset.
 \begin{figure*}[tbp]
 \centering
 \resizebox{0.9\textwidth}{!} {
\includegraphics[]{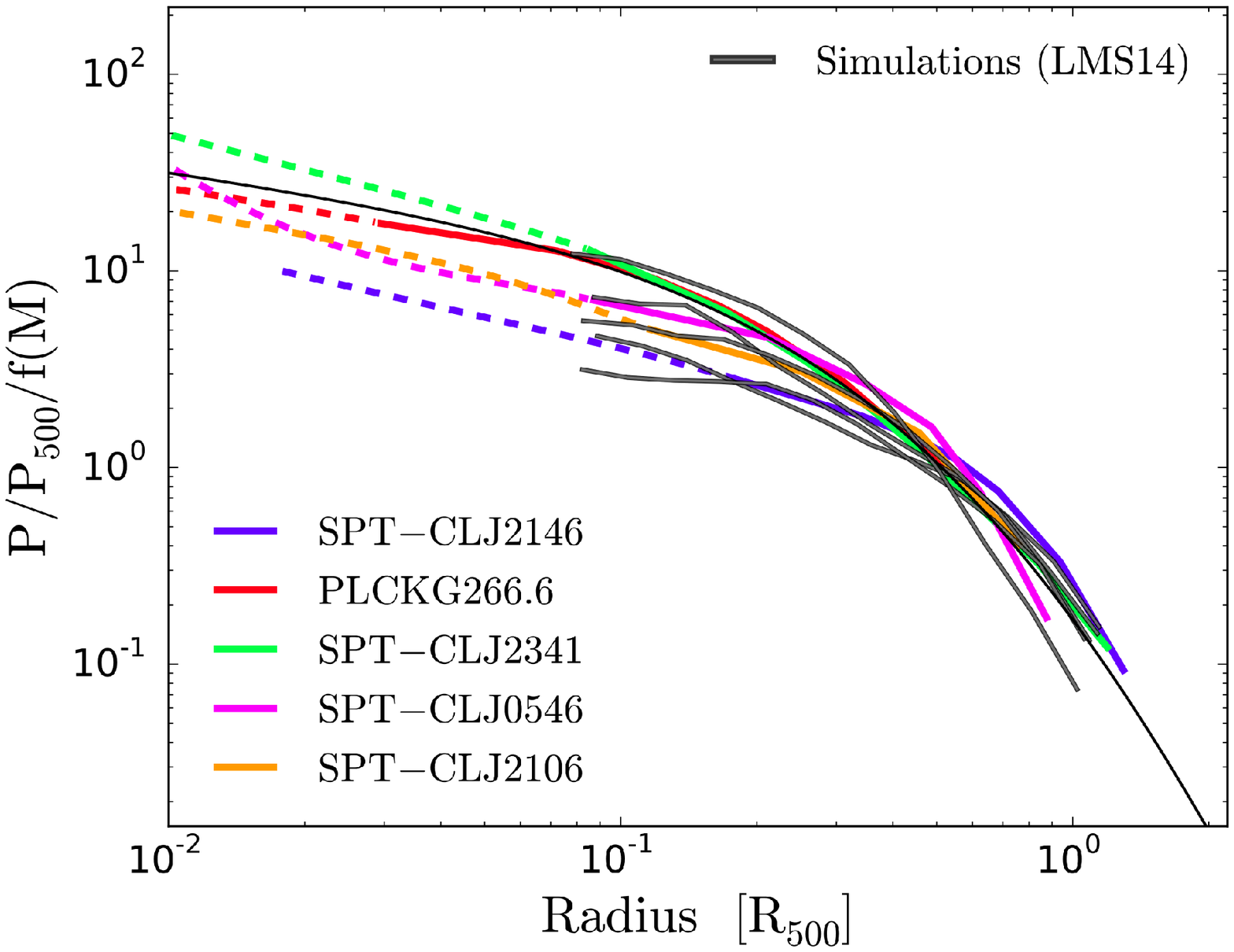}
\hspace{0.05\textwidth}
\includegraphics[]{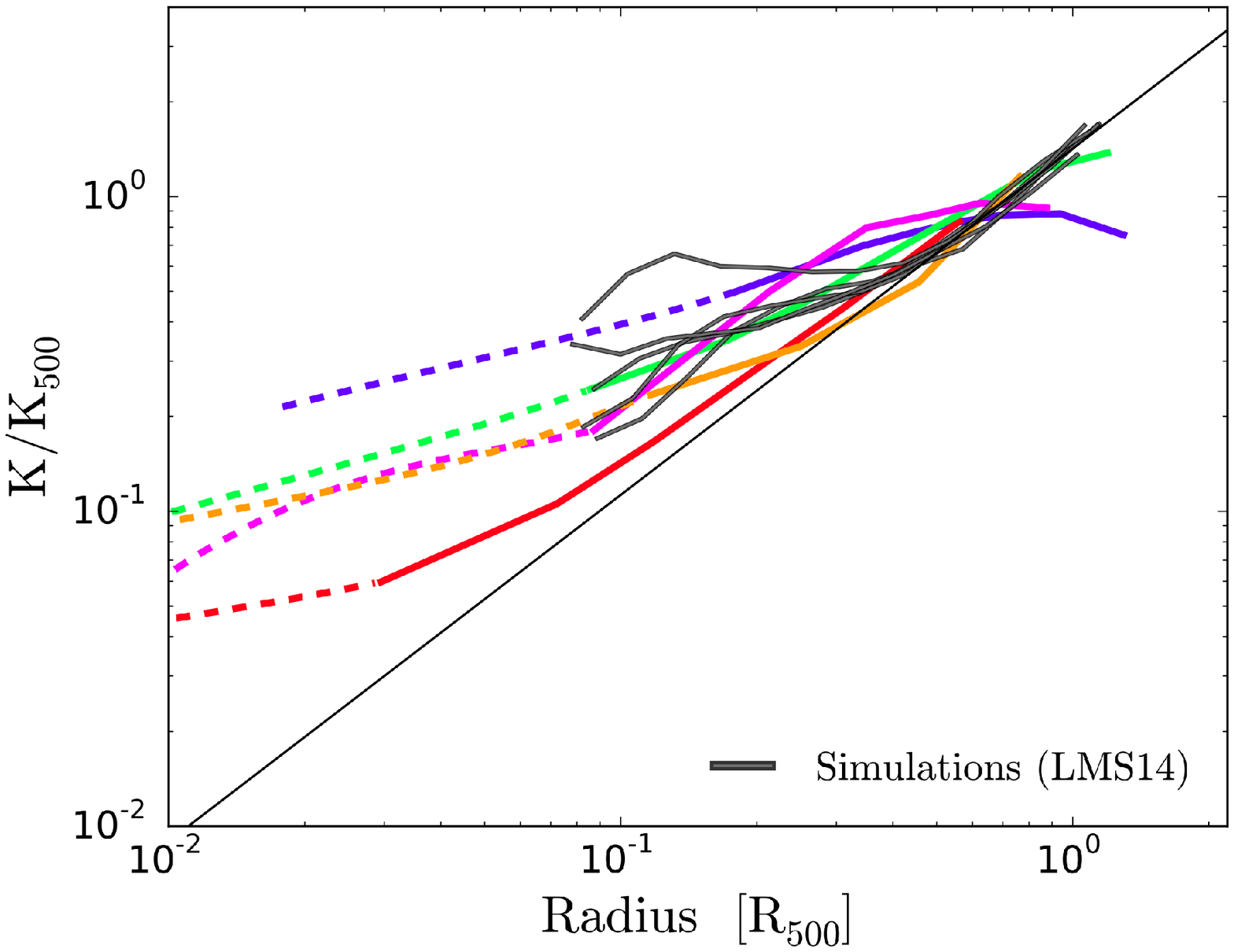}
}

  \caption{\footnotesize{Left panel: scaled pressure profiles, shown using the same legend as in top left panel of \figiac{fig:pressure_entropy_disp}. Grey-black lines are the profiles extracted from the simulated clusters of \citet{amandine2014} (LMS14). Right panel: same as left except for the fact that we show entropy profiles.}}
 \label{fig:simulation_disp}
\end{figure*}

\subsubsection{Entropy profiles}
In the bottom left panel of \figiac{fig:pressure_entropy_disp} we show the entropy profiles scaled by the $K_{500}$ from \citet{pratt2010}, compared to the \rexcess\ $1\sigma$ dispersion envelopes. 
We also show the entropy profile we would expect if the structure formation were driven only by gravitational processes, as computed through simulations in \citet{voit2005all}.
Except for \plck, up to $R \sim 0.4\,R_{500}$ all our of profiles are in excellent agreement with both reference samples. The entropy profile of \plck\ cluster in fact follows closely the gravity-only simulations. This is the most massive cluster in our sample and we expect that in this regime non-gravitational effects are less important; in addition, it is the only cluster that is classified as relaxed. For radii $> 0.5R_{500}$ the profiles show lower entropy with respect to \rexcess, and flatten towards the outskirts. As discussed in \citet{spt2014}, this behaviour may be related to gas clumpiness in the cluster outskirts \citep[see e.g.][]{nagai2011,vazza2013}. This effect boosts X-ray emission of cold gas, cooling temperature profiles in the outskirts and increasing the azimuthally ßaveraged density profiles. 
We also show the entropy scaled values at fixed radii as a function of the mass in the bottom right panel of \figiac{fig:pressure_entropy_disp}. As with \rexcess, we find that in the central part the most massive clusters have smaller entropy. Above $0.7\, R_{500}$, we clearly see the lack of entropy as compared to \rexcess. As for the scaled pressure profiles, there is good agreement in the average dispersion of our sample and that of \rexcess.

As for the density and pressure profiles, we find the same results using BCG centred profiles.

\section{Comparison with \chandra\ stacked analysis}
\label{sec:compxvp}

\citet{spt2014} analysed the redshift evolution of $80$ SPT galaxy clusters observed with \chandra, dividing their sample into two redshift bins, and deriving stacked profiles of temperature, pressure, and entropy centred on the large-scale X-ray centroid.  They classified objects as  cool-core or non-cool-core based on the cuspiness of the density profile, where half of the sample was put into each category by construction. Our morphological classification  is different, being based on the $\langle w \rangle$ dynamical indicator as defined for the \rexcess\ local sample. As a basis for comparison to our work, we  thus  consider the stacked profiles of the full SPT sample in their $0.6 < z < 1.2$ redshift bin.   

The stacked pressure profile from \citet{spt2014} is plotted with open black squares in the top left panel of Fig.~\ref{fig:pressure_entropy_disp}. The agreement in shape is  good, but there is a clear normalisation offset  of $\sim 10-20\%$ with respect to \rexcess\footnote{This result is in contrast to \citeauthor{spt2014}, who found good agreement between their high-redshift subsample and the \rexcess\ profile for $R>0.4\, \Rv$ after correction for cross-calibration differences. From their Fig.~10, we suspect that this may be due to a sign error in the $P/P_{500}$ correction.}  and to the present $z\sim1$ sample.  In comparing the profiles, we have to consider cross-calibration differences in temperature,  as discussed by \citet{spt2014} when comparing their \chandra\ profiles to the \rexcess\ profile obtained from \xmm\ data.  However, the observed offset is unlikely to be due to this difference. The radius $R/\Rv$ scales as $T^{-0.19}$ for a slope of the \MY\  relation of $0.56$, and $P/P_{500} \propto T^{0.63}$. 
Correcting for a maximum difference of $15\%$ in temperature \citep{martino2014} (as assumed by \citeauthor{spt2014}), would translate the curve by $-3\%$ and $+10\%$ along the X- and Y-axis, respectively. As the pressure decreases with radius, the net effect is negligible (indeed, it would be null for a logarithmic slope of $-3$). The insensitivity of the scaled pressure profile to temperature calibration differences is also confirmed by the good agreement between local \chandra\ and \xmm\ profiles shown in \citet{arnaud2010}. 
Furthermore, while the $\TX$ values of the present sample are high  (up to $10$\, keV),  the effective cluster temperatures $(1+z)T$  are lower, around $5$\, keV, in a regime where the cross-calibration difference becomes negligible.  We will thus neglect temperature cross-calibration differences in the following. 

As discussed in \citet{spt2014}, the profile shape can be significantly affected by the choice of centre; we note that their analysis uses the large-scale centroid, whereas ours uses the X-ray peak. As illustrated in their Fig.~6, the difference in the density profile can be large even at large radii, where the profiles centred on the centroid lie below those centred on the X-ray peak. It is thus  possible that some of the offset is due to the different choice of centre. We note that we do not see this difference between the BCG- and X-ray-peak centred profiles in our sample: the distance in our case is probably smaller that that between the peak and the large-scale centroid.

The bottom left panel of \figiac{fig:pressure_entropy_disp} compares the entropy profiles. Here the agreement between the entropy profiles of our sample and the stacked result from \citet{spt2014} is good, taking into account the larger dispersion. Both show lower entropy with respect to \rexcess\ at $R> 0.5\,\Rv$, and some evidence of flattening in the outermost radii.

\section{Comparison with numerical simulations}
\label{sec:compsimul}

We now turn to a comparison with cosmological hydrodynamical simulations. We selected the five galaxy clusters in the mass range $[4-6] \times 10^{14} M_{\odot}$ at $z=1$ from the cosmo-OWLS simulations described in \citet{amandine2014}. This suite is a large-volume extension of the OverWhelmingly Large Simulations project \citep[OWLS][]{schaeye2010}, including baryonic physics such as supernova and various levels of AGN feedback, undertaken to help improve the understanding of cluster astrophysics and non-linear structure formation and evolution. 

The profiles extracted from the simulated clusters were centred on the bottom of the potential well. The objects are all disturbed according to the morphological indicator calculated from the displacement between the bottom of the potential well and the centre of the mass.

The gas and total baryon fraction at $\Rv$ are compared to the observed sample in Fig.~\ref{fig:gas_fraction}. Although the mass range is very limited, there is no obvious mass dependence, in agreement with the conclusions of \citet{lebrun2016}. This effect likely occurs because at $z\sim 1$ clusters at fixed mass are denser, hence the energy required to expel gas from the haloes is greater. The baryonic mass fraction thus stays near the universal value (the value computed from \citealt{planck2015} is shown with a yellow rectangle).

The simulated pressure profiles shown in the left-hand panel of Fig.~\ref{fig:simulation_disp} are in excellent  agreement with the observations; in particular, the observed dispersion is very well captured over the full radial range. However, the simulated entropy profiles show much less dispersion, with all simulated profiles falling very close to the result from \citet{voit2005all}  at $R>0.4R_{500}$. Unlike in the observed profiles, there is no hint of flattening at large radii in the simulations. This may again point to the possible influence of clumping on the observationally derived entropy.

\section{Conclusions}\label{sec:conclusions}

We have presented a spatially resolved X-ray spectroscopic analysis of the five most massive ($M_{500} > 5 \times 10^{14}$ M$_{\odot}$), distant ($z>0.9$) clusters thus-far detected in SZ surveys by \planck\ and SPT. All objects were observed by both \xmm\ and \chandra, and we have investigated in detail how best to combine the datasets. Regarding the complementarity between instruments, our main results are as follows:

\begin{itemize}

\item We have proposed a new technique to combine X-ray datasets that allows us to fully exploit the complementarity between the high throughput of \xmm\ and the excellent angular resolution of \chandra. Finding that the instruments are in excellent agreement concerning flux-related measurements, we performed a joint-fit to the density profiles using a modified $\beta$ model. The resulting combined density profile is constrained in the centre by \chandra, and in the outskirts by \xmm.  

\item We investigated in detail how the choice of centre for profile extraction affects the analysis, finding that profiles centred on the X-ray peak are, as expected, more peaked than when centred on the BCG. This difference is significant in the core, $R < 0.1R_{500}$, but it does not affect profile shape or normalisation in the outer parts of the clusters. Temperature profiles are unaffected by the choice of centre owing to the need for larger bin sizes to build and fit a spectrum.

\item We find that a combination of the two instruments is fundamental for a correct determination of the cluster morphology through the centroid-shift parameter $\langle w \rangle$. At these redshifts, the high resolution of \chandra\ is essential to remove contamination from point sources. At the same time, the high throughput of \xmm\ gives increased photon statistics and is critical for the detection of faint emission from substructures at large radii.

\end{itemize}

We then investigated evolution through comparison with the local X-ray selected sample \rexcess and find the following: 

\begin{itemize}

\item Using the centroid shift parameter $\langle w \rangle$ as a morphological indicator, we find that our sample is dominated by morphologically disturbed clusters with $\langle w \rangle > 0.01$. This result is in line with expectations based on theory, where numerical simulations indicate that higher redshift objects are more likely to be undergoing mergers.

\item Scaling the density profiles according to self-similar predictions, we find no clear evidence for evolution.  Four of the five objects have lower central densities than the mean \rexcess\ profile. 
This is expected from the centroid shift results above, and with the expectation that disturbed cluster density profiles are flatter in the central part.

\item We measured the gas-mass profiles and found good agreement with the \rexcess\ sample once the appropriate scaling had been applied. Combining these measurements with stellar mass measurements from the literature allowed us to measure the total baryon-mass and baryon-mass fraction. At $R_{500}$, we found a baryon mass fraction close to that expected from  \citet{planck2015}. There is no mass dependence, although this is likely due to the limited mass range of our sample.

\item We find no clear sign of evolution in the scaled temperature profiles, the mean profile of our sample being in excellent agreement with that of \rexcess.

\item The scaled pressure profiles are also in good agreement with \rexcess, within the dispersion, across the full radial range. We do not find any clear sign of evolution, either in shape or scatter, which is a fundamental result for any SZ survey that uses a detection algorithm that relies on the universality of the pressure profile. 

\item The scaled entropy profiles, as well as the scatter, are in good agreement with the mean \rexcess\ profile and its $1\sigma$ dispersion interior to $0.7\,R_{500}$. However, at larger radii there appears to be lower entropy compared to the local sample, which may be related to increased gas clumping. 

\end{itemize}

We compared our results to the \chandra\ stacked analysis of SPT SZ-selected clusters at $z>0.6$ \citep{spt2014}, finding good agreement in entropy behaviour but a slight normalisation offset ($\sim 10\%$) in pressure. Finally, we compared to the five clusters in the same mass and redshift range as our sample from the cosmo-OWLS simulations \citep{amandine2014}. The shape and scatter of the pressure profiles is well reproduced by the simulations. However, while the central entropy of the simulated clusters approximates the scatter in the observed profiles, beyond $R \gtrsim 0.5\,R_{500}$ the simulated entropy profiles exhibit remarkably little scatter.

Overall, our results illustrate the benefit of spatially resolved analysis of individual objects at high-redshift. This approach is fundamental to infer the statistical properties of the profiles, and in particular the dispersion around the mean. The current sample of five objects already gives a strong constraint on the mean profile and a first idea of its scatter. However, the latter is very sensitive to the number of clusters and to the presence of outliers. Deep observations of a larger number of objects will allow us to place quantitative constraints on the profile scatter. In parallel, for better comparison with theory, numerical simulations with sufficient volume to generate similarly high-mass, high-redshift samples are needed. 

\begin{acknowledgements} 
The authors would like to thank Amandine Le Brun and Ian McCarthy for providing the simulation data in electronic format, and Paula Tarr\'io for helpful comments and suggestions. The authors further thank the anonymous referee for constructive
comments.
This work is based in part on observations made with the Spitzer Space Telescope, which is operated by the Jet Propulsion Laboratory, California Institute of Technology under a contract with NASA. Based on observations made with the NASA/ESA Hubble Space Telescope, obtained from the data archive at the Space Telescope Science Institute. STScI is operated by the Association of Universities for Research in Astronomy, Inc. under NASA contract NAS 5-26555.
The scientific results reported in this article are based on data obtained from the \chandra\ Data Archive and observations obtained with \xmm , an ESA science mission with instruments and contributions directly funded by ESA Member States and NASA.
 The research leading to these results has received funding from the European  Research  Council  under  the  European  Union’s  Seventh  Framework
Programme (FP72007-2013) ERC grant agreement no 340519. PM acknowledges funding support from NASA grant GO2-13153X.
 \end{acknowledgements}

\bibliographystyle{aa}
\bibliography{lib_articoli}

\appendix
\section{Gallery}\label{appx:a1}
\begin{figure*}[!ht]
 \begin{center}
\includegraphics[width=0.49\textwidth]{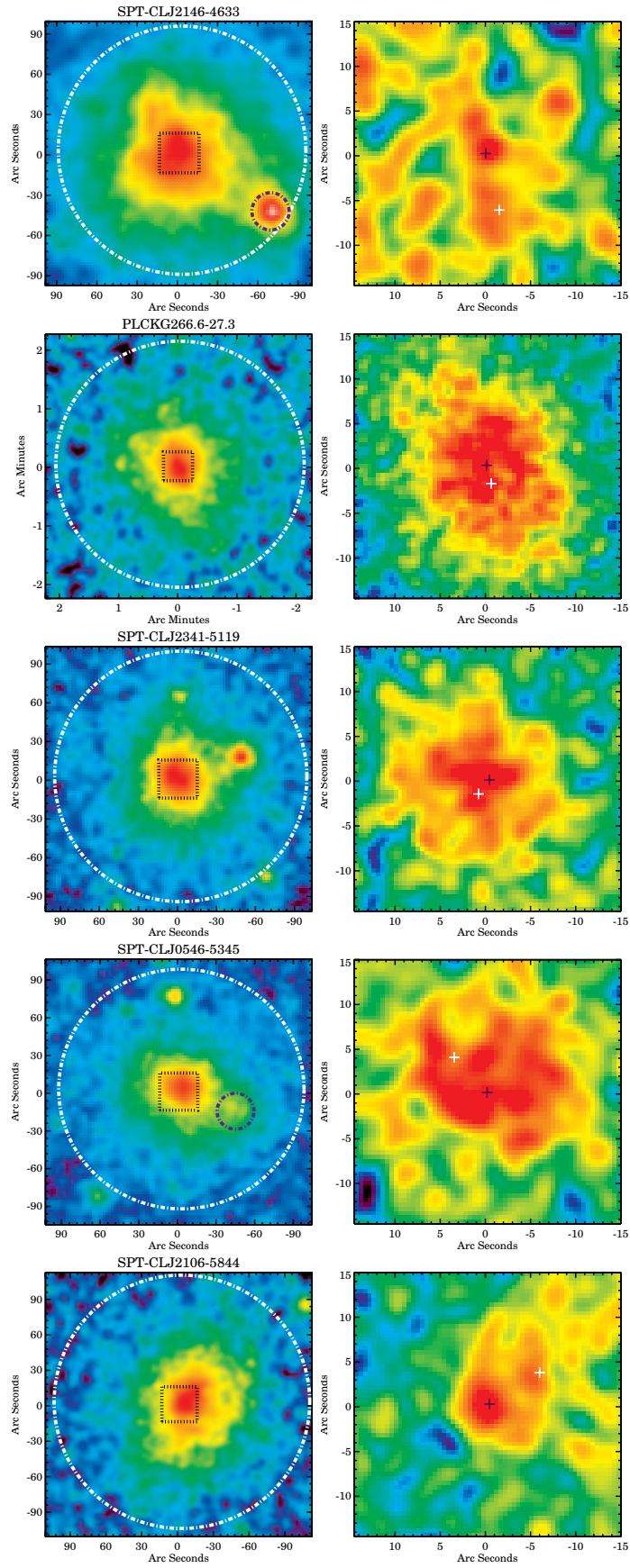} 
  \end{center}
 \caption{\footnotesize{Left column: \xmm\ images in the $[0.3-2]$ keV band. Each image is smoothed to enhance the emission on large scale. The white dotted circle represents $R_{500}$ and the black dotted box represents the field of view of the \chandra\ image on the right column. The blue dotted circle in the SPT-CLJ 2146-4632 and SPT-CLJ 0546-5345 images highlights the substructures. Right column: \chandra\ images in the $[0.5-2.5]$ keV band. Black and white crosses identify the X-ray peak and the BCG position, respectively.}}
 \label{fig:gallery}
\end{figure*}

In \figiac{fig:gallery} we show the \xmm\ and \chandra\ images of the clusters in our sample in the left and right column, respectively. In the left-hand images, the white circle is $R_{500}$ and the black box is the field of view of the \chandra\ image on the right. The substructures in the SPT-CLJ 2146-4632 and SPT-CLJ 0546-5345 \xmm\ images are highlighted with blue dotted circles. It is worth noting that all the other sources not highlighted are consistent with being point sources. They are removed from the calculation of $\langle w \rangle$. 
For the sake of clarity we note that in the SPT-CLJ 2146-4632 \xmm\ image a point source resolved by \chandra\ superimposes on the substructure diffuse emission.

\section{BCG profiles }\label{appx:a2}

 \begin{figure*}[t]
 \begin{center}
\includegraphics[width=1.02\textwidth]{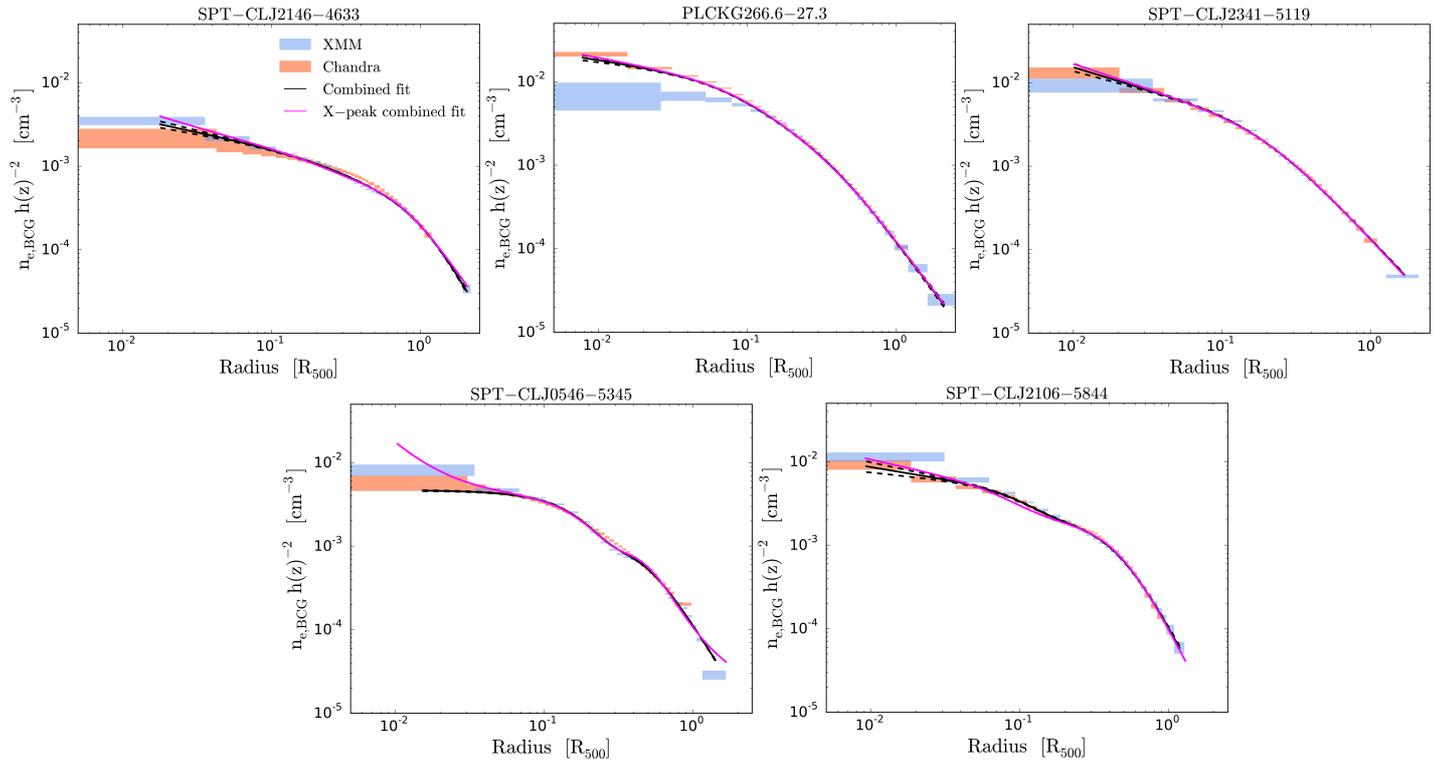}  
  \end{center}
 \caption{\footnotesize{Same as \figiac{fig:hyb_density_xpeak} except that deprojected density profiles are here shown centred on the BCG.}}
 \label{fig:hyb_density_bcg}
\end{figure*}

\end{document}